\documentclass{iopjournal}

\usepackage{enumitem}
\usepackage{url}
\usepackage{amsmath}
\usepackage{amsfonts}
\usepackage{amsthm}
\usepackage{amssymb}
\usepackage{caption}
\usepackage{subcaption}
\usepackage{float}
\usepackage{siunitx}
\usepackage{comment}
\usepackage{booktabs}
\usepackage{multirow}
\usepackage{algorithm}
\usepackage{algpseudocode}
\usepackage{pifont}
\providecommand{\orcidlink}[1]{}
\newcommand{\cmark}{\ding{51}}
\newcommand{\xmark}{\ding{55}}

\newcommand{\tr}[0]{\mathrm{tr}}

\newcommand{\lem}[1]{\hyperref[lem:#1]{Lemma~\ref*{lem:#1}}}

\begin{document}

\articletype{Paper}

\title{Quantum Adaptive Self-Attention for Quantum Transformer Models}

\author{Chi-Sheng Chen$^1$\,\orcidlink{0000-0000-0000-0000} and En-Jui Kuo$^{1,*}$\,\orcidlink{0000-0000-0000-0000}}

\affil{$^1$Department of Electrophysics, National Yang Ming Chiao Tung University, Hsinchu, Taiwan}

\affil{$^*$Author to whom any correspondence should be addressed.}

\email{ejkuo@nycu.edu.tw}


\begin{abstract}
A recurring weakness in quantum machine learning (QML) is that reported ``quantum advantages'' are seldom tested against a \emph{capacity-matched} classical control, leaving it unclear whether a gain comes from the quantum substrate or from the architectural change that accompanies it. Our primary contribution is methodological: a protocol for attributing such gains honestly---a capacity-matched classical bottleneck of identical parameter budget, transparent reporting of where quantum does \emph{not} help, and validation on real quantum hardware---which we develop and apply through a concrete case study. That case study is Quantum Adaptive Self-Attention (QASA), a hybrid Transformer that replaces the value projection of a \emph{single} encoder layer with a 36-parameter parameterized quantum circuit (PQC), keeping all other layers classical. Across nine synthetic benchmarks and the real-world ETTh1 dataset, QASA improves on a full-capacity classical Transformer for chaotic and trend-dominated signals. To ask whether this is a genuinely \emph{quantum} effect, we introduce a control rarely applied in quantum machine learning---a capacity-matched classical bottleneck with the same parameter budget---and find that it matches the PQC on the error metrics. The gain is therefore attributable to the low-rank value-projection \emph{bottleneck} (an \emph{architectural parsimony} principle), not to quantumness; adding further quantum layers only degrades performance and trainability. We accordingly position the quantum layer not as a source of accuracy advantage but as a \emph{competitive} instantiation of this principle: its low-rank compression onto the signal's intrinsic dimensionality is matched by a classical bottleneck, so the gain is architectural rather than quantum. The quantum layer's distinguishing features are physical---a high circuit entanglement (Meyer--Wallach $Q=0.981$ with only 27 CNOTs, which we report as a circuit-level property) and, most concretely, NISQ deployability, which we verify by executing the trained model on a real IBM Quantum processor (one-step prediction within $7.5\%$ of noiseless simulation on the quantum-favoured task and $25\%$, with overlapping error bars, on a classical-favoured control; no error mitigation). Relative to other quantum sequence models (QLSTM, QnnFormer), QASA is moreover the most resource-efficient of the three: it reaches competitive accuracy with the fewest quantum parameters (36 vs.\ 90--128) and the strongest entanglement-per-gate ($Q=0.981$ with 27 CNOTs vs.\ QLSTM's 56), and retains trainable gradients that deeper circuits lose. We argue that capacity-matched baselines and honest reporting of where quantum does \emph{not} help are prerequisites for credible quantum-machine-learning claims.
\end{abstract}

\section{Introduction}
Transformer architectures~\cite{vaswani2017attention} have become the backbone of modern deep learning, powering state-of-the-art models across natural language processing, computer vision, and sequential prediction. Their core strength lies in the self-attention mechanism, which dynamically models long-range dependencies. However, this comes at a significant computational cost---quadratic in sequence length---motivating the search for alternative attention paradigms.

Quantum computing offers a compelling candidate: parameterized quantum circuits (PQCs) operate in exponentially large Hilbert spaces and can represent correlations that are classically intractable to express compactly~\cite{havlicek2019supervised}. Recent work has explored quantum-enhanced sequence models, including quantum LSTMs~\cite{chen2022quantum} and quantum Transformer variants~\cite{cai2024qnnformer}, which replace classical components wholesale with variational quantum circuits. However, a fundamental question remains unanswered: \emph{how much quantum integration is actually beneficial, and where should it be placed?}

In this work, we propose Quantum Adaptive Self-Attention (QASA), a hybrid architecture designed around a principle of \textbf{architectural parsimony}---the idea that the most one can gain from a quantum layer comes not from maximizing quantum resources, but from placing minimal quantum computation at the optimal position. QASA retains $N{-}1$ classical Transformer encoder layers and replaces only the value projection in the \emph{final} layer with a PQC, using just 36 trainable quantum parameters. This stands in contrast to approaches such as QLSTM (128 quantum parameters) and QnnFormer (90 quantum parameters), which distribute quantum computation across the entire model.

Our study is organised around a methodological commitment that, we argue, is too often missing from the quantum-machine-learning literature: we ask not merely whether a quantum model outperforms prior \emph{quantum} baselines, but whether the quantum component outperforms a \emph{capacity-matched classical control}. Concretely, we compare QASA's $36$-parameter quantum value map against a classical low-rank value map of the same parameter budget, holding every other component fixed. This control is decisive, and its outcome shapes our claims: we find---and report transparently---that the performance gain over a full-capacity Transformer is driven by the low-rank value-projection \emph{bottleneck}, which the classical control exploits equally well. Rather than weakening the paper, this sharpens it: it lets us cleanly separate \emph{what the architecture buys} (parsimony, including the low-rank compression---which we show a classical bottleneck achieves equally well) from \emph{what the quantum substrate uniquely buys} (entanglement structure and---demonstrated on real IBM Quantum hardware---NISQ deployability), and it offers the community a template for substantiating, or appropriately qualifying, quantum-advantage claims.

Our systematic evaluation across nine synthetic benchmarks, one real-world dataset, two quantum baselines (QLSTM and QnnFormer) alongside a classical Transformer, a capacity-matched classical bottleneck control, and a comprehensive ablation study yields the following key contributions:

\subsection{Contributions}
Our contributions are primarily \emph{methodological}, with QASA as the case study that demonstrates them:
\begin{itemize}
    \item \textbf{A capacity-matched methodology for QML claims (primary contribution).} We argue that a quantum component should be compared not only to prior \emph{quantum} baselines but to a \emph{capacity-matched classical control}---a low-rank classical value map of the same parameter budget, with every other component held fixed---and we carry this control, together with circuit-level characterisation, multi-channel noise tests, and real-hardware execution, through a complete study. To our knowledge this is the most thoroughly capacity-controlled comparison in the quantum-transformer literature, and we offer it as a reusable template for substantiating, or appropriately qualifying, quantum-advantage claims.

    \item \textbf{The gain is the bottleneck, not quantumness---and we say so.} Applying this control (Table~\ref{tab:bottleneck}), we show that QASA's advantage over a full-capacity classical Transformer comes from the low-rank value-projection \emph{bottleneck structure}: a $40$-parameter classical map matches the $36$-parameter PQC on the error metrics (and beats it on clean periodic signals) \emph{and} compresses the representation comparably. We therefore frame the quantum layer as a \emph{competitive} realisation of architectural parsimony, not a source of error-metric or compression advantage; its distinguishing value is physical---high circuit entanglement reported as a property, barren-plateau-aware trainability, and NISQ deployability verified on a real IBM Quantum processor. We view this candour as itself a contribution to a literature where such baselines are frequently omitted.

    \item \textbf{Architectural parsimony principle.} We demonstrate that a single quantum-enhanced layer is competitive with architectures using 2--4$\times$ more quantum parameters. Our ablation study indicates that quantum layer \emph{position} matters more than \emph{count}: increasing quantum layers from one to four degrades performance on most tasks, while a single well-placed layer matches or outperforms all multi-layer configurations.

    \item \textbf{Task-conditional taxonomy.} We provide empirical evidence that a parsimonious low-rank bottleneck (classical or quantum) benefits specific signal classes---chaotic dynamics and smooth trends (with noisy oscillations a statistical tie)---while unconstrained, full-capacity classical Transformers suit clean periodic signals and sharp discontinuities. This taxonomy offers practitioners a principled guide for when a minimal bottleneck (in either form) is warranted.

    \item \textbf{An efficient, trainable, hardware-validated quantum realisation.} Where the comparison \emph{is} favourable---against other quantum sequence models and on physical/deployment criteria---QASA holds clear, honestly-scoped advantages. Among the quantum baselines it is the most resource-efficient: it reaches competitive accuracy (equal-or-better MSE on $7/9$ tasks vs.\ QLSTM) with the \emph{fewest} quantum parameters ($36$ vs.\ QnnFormer's $90$ and QLSTM's $128$) and generates the \emph{strongest} entanglement ($Q=0.981$) using only $27$ CNOTs (vs.\ QLSTM's $56$). It is trainability-aware: a single shallow quantum encoder layer keeps gradients trainable where deeper circuits decay toward a barren plateau ($146\times$ larger gradient variance than the 4-layer variant at $12$ qubits). And it is NISQ-ready: degradation is bounded across depolarizing, amplitude-damping, and bit-flip channels, and we execute the trained model on a real IBM Quantum processor with prediction quality preserved to within $7.5\%$ of noiseless simulation, without error mitigation.

\end{itemize}

\begin{figure}
    \centering
    \includegraphics[width=0.9\textwidth]{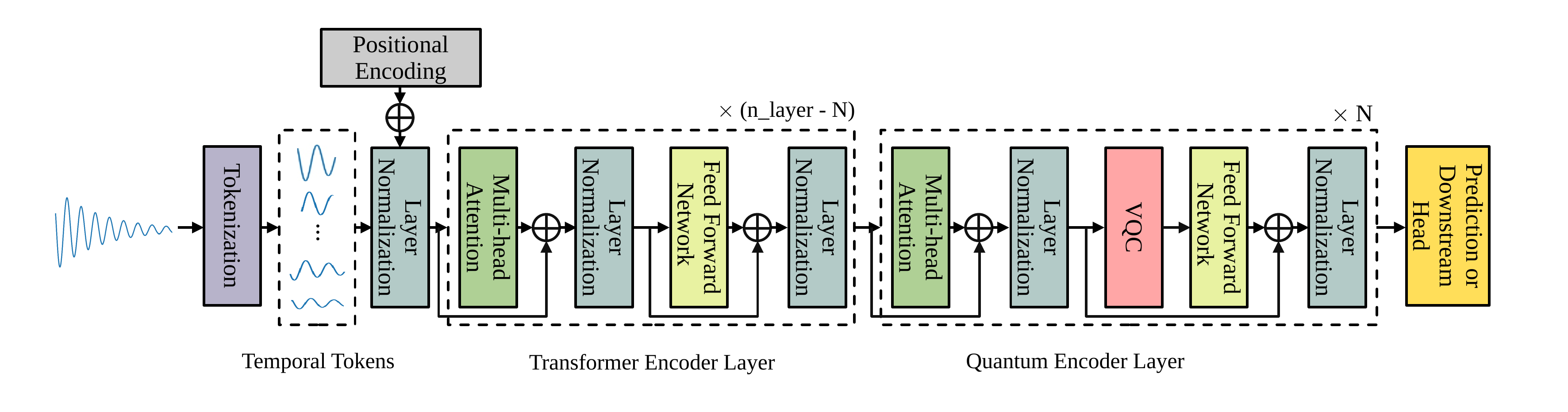}
    \caption{Overview of the QASA architecture. The model consists of $N{-}1$ standard Transformer encoder layers followed by one Quantum Encoder Layer. The Quantum Encoder Layer replaces the value projection in self-attention with a parameterized quantum circuit (PQC) operating on 8 qubits with 4 variational layers, followed by a classical feedforward network with residual connections.}
    \label{fig:qasa}
\end{figure}

\section{Related work}
\subsection{Deep Learning on Time-Series Data}
Deep learning has significantly advanced the modeling and understanding of time-series data across various domains. Recurrent Neural Networks (RNNs) \cite{schuster1997bidirectional}, particularly Long Short-Term Memory (LSTM) networks \cite{hochreiter1997long} and Gated Recurrent Units (GRUs) \cite{chung2014empirical}, have been widely adopted due to their ability to capture long-term temporal dependencies. However, their sequential nature limits parallelization and increases training time. To address this, temporal convolutional networks (TCNs) \cite{lea2017temporal} and Transformer-based architectures \cite{vaswani2017attention} have demonstrated superior performance by enabling parallel processing and better long-range dependency modeling.

In industrial and scientific applications, Transformer variants such as the Temporal Fusion Transformer (TFT) \cite{lim2021temporal} and Informer \cite{zhou2021informer} have been applied to multi-horizon prediction tasks with irregular time intervals and exogenous variables. Similar time-series learning techniques have been applied successfully in molecular dynamics and open quantum systems \cite{tsai2022path, tsai2020learning}.

Despite these advancements, challenges remain in modeling noisy, sparse, or irregularly-sampled sequences, motivating the development of hybrid models that integrate domain-specific priors with general-purpose deep learning architectures.

\subsection{Quantum Deep Learning on Time-Series Data}
While classical deep learning has achieved remarkable success in time-series analysis, it often requires extensive computational resources and large-scale data to generalize effectively. With the advent of quantum computing, Quantum Deep Learning (QDL) \cite{biamonte2017quantum} has emerged as a promising paradigm for modeling complex temporal patterns, offering theoretical advantages in expressivity and optimization through quantum entanglement and superposition.

In the context of time-series data, several quantum-inspired architectures have been proposed. Variational Quantum Circuits (VQCs) \cite{gohel2024quantum} and Quantum Recurrent Neural Networks (QRNNs) \cite{takaki2021learning} have been introduced to capture temporal dependencies using parameterized quantum gates and recurrent structures. A particularly notable development is the Quantum Long Short-Term Memory (QLSTM) network \cite{chen2022quantum}, which integrates quantum circuits into the gating mechanisms of classical LSTMs. Recent comprehensive benchmarking studies \cite{fellner2025quantum_benchmark, chen2025benchmarking_sequential} have compared VQC-based regression against classical GRU and LSTM models, finding that while classical models often lead in raw accuracy, VQCs exhibit higher robustness to noise and better generalization on small-scale datasets. Quantum-enhanced channel mixing has also been explored in state-space models for temporal forecasting~\cite{chen2025quantum_rwkv}, while quantum temporal convolutional networks have been applied to cross-sectional equity return prediction~\cite{chen2025quantum_tcn}.

Recent work has further expanded quantum approaches for temporal data. Tang et al.~\cite{tang2026qnnformer} proposed QnnFormer, a multi-attention mechanism leveraging quantum circuits for long-term forecasting, achieving significant error reductions on benchmark datasets. Chakraborty and Heintz~\cite{chakraborty2025qcaapatch} integrated hybrid quantum-classical attention into patch-based transformers for multivariate forecasting. Chittoor et al.~\cite{chittoor2024qultsf} developed QuLTSF for long-term forecasting with quantum neural networks, demonstrating advantages in convergence stability for non-stationary data. Laskar and Goel~\cite{laskar2025shallow} investigated the impact of different entanglement topologies in shallow circuits for time-series prediction on NISQ hardware. In finance, quantum-enhanced architectures have been investigated for demand prediction and volatility modeling \cite{paquet2022quantumleap}.

Beyond time series, quantum machine learning has been applied to sequential data in other domains, including EEG encoding~\cite{chen2024qeegnet, chen2025exploring_qeegnet} and multimodal contrastive learning~\cite{chen2024quantum_multimodal}. Chen and Kuo~\cite{chen2025unraveling} demonstrated that Transformer-assisted learning can effectively capture Lindblad dynamics in open quantum systems, establishing a direct connection between Transformer architectures and quantum physics. Despite these advancements, research on Quantum Transformers for time-series data remains limited. Most existing studies focus on Quantum Visual Transformers (QViT) \cite{cherrat2022quantum} or fully quantum architectures such as Quixer \cite{khatri2024quixer} in sequence modeling, leaving a gap in understanding how hybrid quantum-classical self-attention mechanisms can be practically deployed for temporal forecasting. Addressing this gap is the key focus of this work.

\section{Methodology}
\begin{figure}
    \centering
    \includegraphics[width=0.9\textwidth]{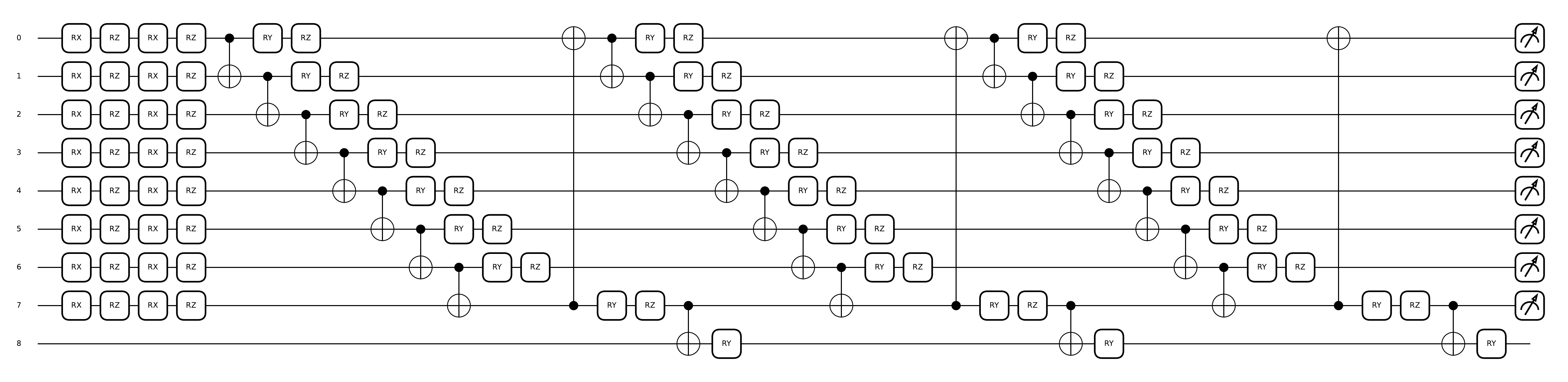}
    \caption{The parameterized quantum circuit (PQC) used in QASA's QuantumLayer. The circuit operates on $n{=}8$ data qubits plus one ancilla qubit. Input features are encoded \emph{once} via $R_X$ and $R_Z$ rotations on each qubit; this is followed by trainable variational layers (single-qubit $R_X$, $R_Y$, $R_Z$ rotations; see Eq.~(\ref{eq:rotation_defs})) with ring-topology CNOT entanglement and an ancilla coupling. The deployed circuit has $36$ trainable angles and $27$ CNOTs in total (exact counting convention in Table~\ref{tab:circuit_analysis}). The output is obtained by measuring $\langle Z_i \rangle$ on all data qubits.}
    \label{fig:qasa_qnn}
\end{figure}

\begin{table}[h]
\centering
\caption{Quantum Adaptive Self-Attention (QASA) Transformer Architecture. Each row lists a layer with its operation and input/output tensor shapes.
$L$ denotes the input sequence length (e.g., $L{=}50$),
$d$ is the hidden feature dimension (e.g., $d{=}256$),
and $n$ is the number of qubits in the quantum circuit (e.g., $n{=}8$). 'Attn' is an abbreviation for 'attention'.}
\label{tab:model-architecture}
\begin{tabular}{l l c c }
\hline
\textbf{Layer} & \textbf{Operation} & \textbf{Input Shape} & \textbf{Output Shape} \\
\hline
Input & Raw sequence & $(L, 1)$ & $(L, 1)$ \\
Linear Embedding & Linear + LayerNorm & $(L, 1)$ & $(L, d)$ \\
Positional Encoding & Add sinusoidal PE & $(L, d)$ & $(L, d)$ \\
Transformer Layer $\times (N{-}1)$ & Multihead Attn + FFN & $(L, d)$ & $(L, d)$ \\
Quantum Encoder Layer & \begin{tabular}[c]{@{}l@{}}Self-Attn + QNN + FFN\end{tabular} & $(L, d)$ & $(L, d)$ \\
\quad QuantumLayer (QNN) & \begin{tabular}[c]{@{}l@{}}Linear $\rightarrow \mathbb{R}^n$ \\ PQC $\rightarrow \mathbb{R}^n$ \\ Linear $\rightarrow \mathbb{R}^d$ + Residual\end{tabular} & $(L, d)$ & $(L, d)$ \\
Final Linear & Extract $h[L]$ and project & $(d)$ & $(1)$ \\
\hline
\end{tabular}
\end{table}

\subsection{Overview}

We propose a hybrid quantum-classical Transformer model tailored for sequential data forecasting. The architecture is designed to capture temporal dependencies through classical attention mechanisms, while integrating a parameterized quantum circuit (PQC) to enhance the model's expressiveness and representation power.

Given an input sequence $x \in \mathbb{R}^{L \times 1}$ of length $L$, our model predicts a scalar target $\hat{y} \in \mathbb{R}$ corresponding to the next value in the sequence.

\subsection{Embedding and Positional Encoding}

The input sequence is first projected into a high-dimensional space using a linear layer, followed by layer normalization:
\begin{equation}
h_0 = \text{LayerNorm}(W_e x + b_e), \quad h_0 \in \mathbb{R}^{L \times d},
\end{equation}
where $d$ is the hidden dimension. We then apply fixed sinusoidal positional encoding to inject temporal order information:
\begin{equation}
h_0 \leftarrow h_0 + PE,
\end{equation}
where $PE$ denotes the positional encoding matrix.

\subsection{Transformer Encoder Layers}

The encoder consists of $N$ layers, where the first $N-1$ layers are standard Transformer encoder layers defined as:
\begin{equation}
h_i = \text{Transformer}(h_{i-1}), \quad i = 1, \dots, N-1.
\end{equation}

The Transformer layer consists of two main sublayers: a multi-head self-attention mechanism and a position-wise feed-forward network (FFN). Each sublayer is wrapped with a residual connection and layer normalization. Formally, the computation of the $i$-th Transformer layer can be described as follows:
\begin{align}
z_i &= \text{LayerNorm}\left(h_{i-1} + \text{MultiHeadSelfAttention}(h_{i-1})\right), \\
h_i &= \text{LayerNorm}\left(z_i + \text{FFN}(z_i)\right).
\end{align}

The multi-head self-attention mechanism is defined as:
\begin{equation}
\text{MultiHeadSelfAttention}(X) = \text{Concat}(head_1, \dots, head_H)W^O,
\end{equation}
where each attention head is computed as:
\begin{equation}
head_j = \text{Attention}(XW_j^Q, XW_j^K, XW_j^V), \quad j = 1, \dots, H.
\end{equation}

In the self-attention mechanism, each input vector is linearly projected into three different spaces to form the \textbf{query} ($Q$), \textbf{key} ($K$), and \textbf{value} ($V$) matrices:
\begin{equation}\label{eq:at}
Q = XW^Q, \quad K = XW^K, \quad V = XW^V,
\end{equation}
where $X \in \mathbb{R}^{T \times d_{\text{model}}}$ is the input sequence (with $T$ tokens), and $W^Q, W^K, W^V \in \mathbb{R}^{d_{\text{model}} \times d_k}$ are learnable projection matrices.

The roles of $Q$, $K$, and $V$ are as follows:
\begin{itemize}
  \item \textbf{Query ($Q$)}: Represents the current token's request for information.
  \item \textbf{Key ($K$)}: Represents the "content" or identity of each token in the sequence.
  \item \textbf{Value ($V$)}: Represents the actual information to be retrieved.
\end{itemize}

The attention mechanism computes a similarity score between each query and all keys:
\begin{equation}
\text{Attention}(Q, K, V) = \text{softmax}\left( \frac{QK^\top}{\sqrt{d_k}} \right) V.
\end{equation}
This allows the model to retrieve contextually relevant information by weighting each value according to the query-key similarity.

The feed-forward network (FFN) is applied to each position separately and identically:
\begin{equation}
\text{FFN}(x) = \max(0, xW_1 + b_1)W_2 + b_2,
\end{equation}
where $W_1, W_2, b_1, b_2$ are learnable parameters.

Each Transformer encoder layer includes multi-head self-attention, residual connections, and feedforward networks with GELU activation (the block is exactly Eqs.~(\ref{eq:at})ff.\ above; we do not restate it).

\subsection{Quantum Encoder Layer}

The final encoder block is a quantum-enhanced encoder layer. It begins with the same multi-head self-attention as the classical layers (Eqs.~(\ref{eq:at})ff.), $a=\mathrm{MultiHeadAttention}(h_{N-1})$, followed by a residual LayerNorm:
\begin{equation}
h' = \text{LayerNorm}(h_{N-1} + a).
\end{equation}

Each token vector $h'_i \in \mathbb{R}^d$ is passed through a quantum layer. The quantum layer first linearly projects the input to a lower-dimensional space suitable for quantum encoding:
\begin{equation}
h_q = \tanh(W_q h'_i),
\end{equation}
where $h_q \in \mathbb{R}^{n}$ and $n$ is the number of qubits.

The vector $h_q$ is then passed as input to a parameterized quantum circuit (PQC), which is defined over $n + 1$ qubits and $L_q$ layers of unitary operations. The PQC encodes the input with a single layer of $R_X$ and $R_Z$ gates, applies trainable single-qubit rotations across its $L_q$ variational layers, and introduces entanglement using CNOT and $R_Y$ gates. Throughout this paper, the single-qubit rotation gates are defined as
\begin{equation}
\label{eq:rotation_defs}
R_X(\theta) = e^{-i\theta X/2}, \quad R_Y(\theta) = e^{-i\theta Y/2}, \quad R_Z(\theta) = e^{-i\theta Z/2},
\end{equation}
where $X, Y, Z$ are the Pauli matrices. The final output is computed as the expectation values of Pauli-$Z$ operators on the first $n$ qubits:
\begin{equation}
\text{QC}(h_q) = \left[ \langle Z_j \rangle \right]_{j=1}^{n}.
\end{equation}

The quantum output is then projected back to the original dimension and added residually:
\begin{equation}
z_i = h'_i + W_o \cdot \text{QC}(h_q + t),
\end{equation}
where $t$ is the input sequence length $L$---a single scalar, identical for all tokens of a given sequence---added to $h_q$ as a global temporal-context conditioning signal before the rotation gates.

Finally, a feedforward network with GELU activation is applied, followed by layer normalization:
\begin{equation}
h_N = \text{LayerNorm}(z + \text{FFN}(z)).
\end{equation}

\paragraph{QuantumLayer: Residual Quantum Projection with Conditional Encoding.}

To enrich the representational capacity of the Transformer encoder, we introduce a novel \texttt{QuantumLayer} module that integrates parameterized quantum circuits (PQC) into a residual learning structure. Each token embedding $x \in \mathbb{R}^d$ is first projected into a quantum-compatible latent space $\mathbb{R}^n$, where $n$ denotes the number of qubits:

\begin{equation}
h_q = \tanh(W_q x),
\end{equation}

where $W_q \in \mathbb{R}^{n \times d}$ is a learnable projection matrix, and (consistently with Eq.~(\ref{eq:qlayer_out}) and the earlier overview) the scalar sequence-length encoding $t$ is added to $h_q$ \emph{at the PQC input}, i.e.\ the circuit receives $h_q + t$, to condition the quantum processing on global temporal context. We note that with a fixed context length ($L{=}20$ throughout the main benchmarks) $t$ is a constant shared by all tokens and all sequences, so it reduces to a fixed global offset on the encoding angles and its value in this setting is limited; it would act as a genuinely input-dependent conditioning signal only under variable-$L$ training, which we leave to future work.

The resulting vector $h_q$ is fed into a parameterized quantum circuit $\text{QC}(\cdot)$ with $L_q$ layers and $n+1$ qubits, using a single angle-encoding layer followed by trainable variational layers. Each layer of the PQC applies a combination of single-qubit rotations ($R_X$, $R_Y$, $R_Z$) and entangling CNOT gates, allowing the circuit to perform complex non-classical transformations:

\begin{equation}
\text{QC}(h_q) = \left[ \langle Z_j \rangle \right]_{j=1}^{n},
\end{equation}

where $\langle Z_j \rangle$ denotes the expectation value of the Pauli-$Z$ operator measured on qubit $j$.

The output from the quantum circuit is then projected back to the original feature space and added to the input as a residual enhancement:

\begin{equation}
\label{eq:qlayer_out}
\text{QuantumLayer}(x, t) = x + W_o \cdot \text{QC}(h_q {}+ t),
\end{equation}

where $W_o \in \mathbb{R}^{d \times n}$ is a learnable linear projection. This design allows the model to leverage the expressive power of quantum circuits within a fully differentiable classical framework. The quantum circuit acts as a learnable nonlinear operator conditioned on both the feature vector and temporal context, enabling nonlinear transformations of the per-token representation (we do not claim these exceed a capacity-matched classical map; see Section~\ref{sec:datasets}).

\paragraph{Hybrid Classical-Quantum Encoding.}

In our design, we adopt a hybrid encoder composed of $(N{-}1)$ standard Transformer encoder layers followed by a single quantum-enhanced encoder layer. This layered configuration offers a practical and effective trade-off between scalability and expressiveness. Specifically, the classical Transformer blocks serve as powerful hierarchical feature extractors with stable training dynamics, while the quantum encoder layer provides a complementary inductive bias via entangled quantum operations and high-dimensional projection.

By placing the quantum block at the final stage of the encoder, the model benefits from:
\begin{itemize}
    \item \textbf{Stable gradient propagation:} Classical layers handle early-stage representation learning, mitigating the vanishing gradient problem that may arise in purely quantum-based models.
    \item \textbf{Non-classical feature transformation:} The quantum circuit introduces nonlinear, entanglement-based transformations of the per-token features; a capacity-matched classical bottleneck achieves comparable representational compression (Section~\ref{sec:datasets}), so we do not present this as a uniquely quantum capability.
    \item \textbf{Efficient resource usage:} Since only a single quantum layer is used, the model maintains compatibility with current noisy intermediate-scale quantum (NISQ) hardware, and avoids the overhead of full quantum depth throughout the network.
\end{itemize}

This hybrid composition allows the model to leverage the strengths of both classical and quantum computation in a synergistic manner, enabling end-to-end training on standard hardware with enhanced representation capability for sequential learning tasks.

\paragraph{Training via the parameter-shift rule.}
Gradients of the PQC output with respect to its variational parameters are computed using the parameter-shift rule~\cite{mari_bromley_izaac_schuld_killoran_2020}. For any Pauli rotation gate $R_\sigma(\theta) = e^{-i\theta\sigma/2}$ with $\sigma \in \{X, Y, Z\}$, and any observable $\hat{O}$, the partial derivative of the expectation value $\langle \hat{O} \rangle_\theta = \langle \psi(\theta)|\hat{O}|\psi(\theta)\rangle$ satisfies
\begin{equation}
\label{eq:param_shift}
\frac{\partial \langle \hat{O} \rangle_\theta}{\partial \theta}
= \frac{1}{2}\left[\langle \hat{O}\rangle_{\theta + \pi/2} - \langle \hat{O}\rangle_{\theta - \pi/2}\right].
\end{equation}
This identity yields exact (noise-free in the ideal case) gradients through two additional circuit evaluations per parameter, and is compatible with PyTorch's autograd via PennyLane, enabling end-to-end training of the full hybrid model.

\paragraph{QASA forward pass.}
Algorithm~\ref{alg:qasa} summarises the forward pass of QASA for a single input sequence, making the hybrid classical--quantum information flow explicit.

\begin{algorithm}[h]
\caption{QASA forward pass for a single input sequence.}
\label{alg:qasa}
\begin{algorithmic}[1]
\Require Input sequence $x \in \mathbb{R}^{L \times 1}$; hidden dim $d$; encoder depth $N$; qubit count $n$; PQC depth $L_q$
\Ensure Scalar prediction $\hat{y} \in \mathbb{R}$
\State $h_0 \gets \mathrm{LayerNorm}(W_e x + b_e) + \mathrm{PE}$ \Comment{embedding + positional encoding}
\For{$i = 1$ to $N - 1$} \Comment{$N{-}1$ classical Transformer layers}
  \State $z_i \gets \mathrm{LayerNorm}\!\left(h_{i-1} + \mathrm{MultiHeadSelfAttn}(h_{i-1})\right)$
  \State $h_i \gets \mathrm{LayerNorm}\!\left(z_i + \mathrm{FFN}(z_i)\right)$
\EndFor
\State $a \gets \mathrm{MultiHeadSelfAttn}(h_{N-1})$ \Comment{quantum encoder block}
\State $h' \gets \mathrm{LayerNorm}(h_{N-1} + a)$
\For{each token $h'_j$ in the sequence}
  \State $h_q \gets \tanh(W_q h'_j)$ \Comment{classical-to-quantum projection}
  \State $o_j \gets \mathrm{PQC}_\theta(h_q + t)$ where $(o_j)_k = \langle Z_k\rangle$ for $k = 1, \dots, n$ \Comment{$t$: sequence-length conditioning; Eq.~\ref{eq:rotation_defs}}
  \State $z_j \gets h'_j + W_o o_j$ \Comment{residual quantum-to-classical projection}
\EndFor
\State $h_N \gets \mathrm{LayerNorm}(z + \mathrm{FFN}(z))$
\State $\hat{y} \gets W_{\mathrm{out}} h_N[L] + b_{\mathrm{out}}$ \Comment{read out last token}
\State \Return $\hat{y}$
\end{algorithmic}
\end{algorithm}

\begin{figure}[htbp]
  \centering

  \begin{subfigure}[b]{0.3\textwidth}
    \includegraphics[width=\textwidth]{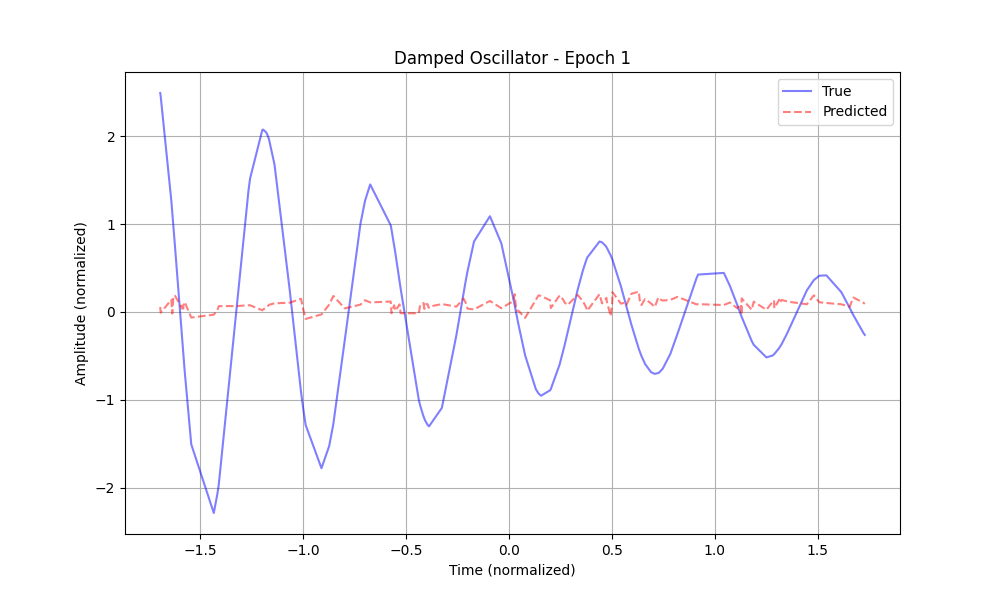}
    \caption{Traditional Transformer - Epoch 1}
  \end{subfigure}
  \begin{subfigure}[b]{0.3\textwidth}
    \includegraphics[width=\textwidth]{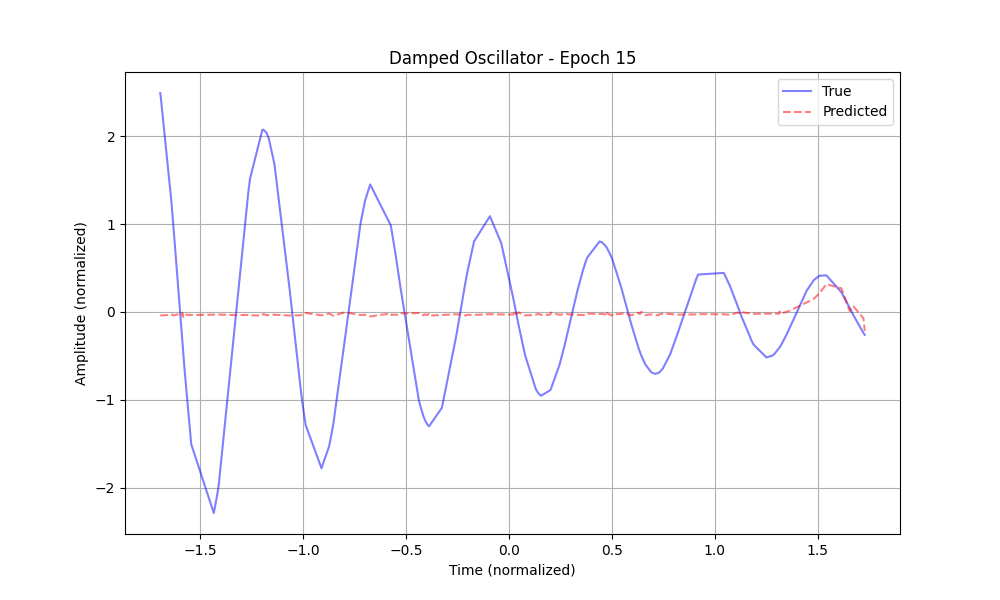}
    \caption{Traditional Transformer - Epoch 15}
  \end{subfigure}
  \begin{subfigure}[b]{0.3\textwidth}
    \includegraphics[width=\textwidth]{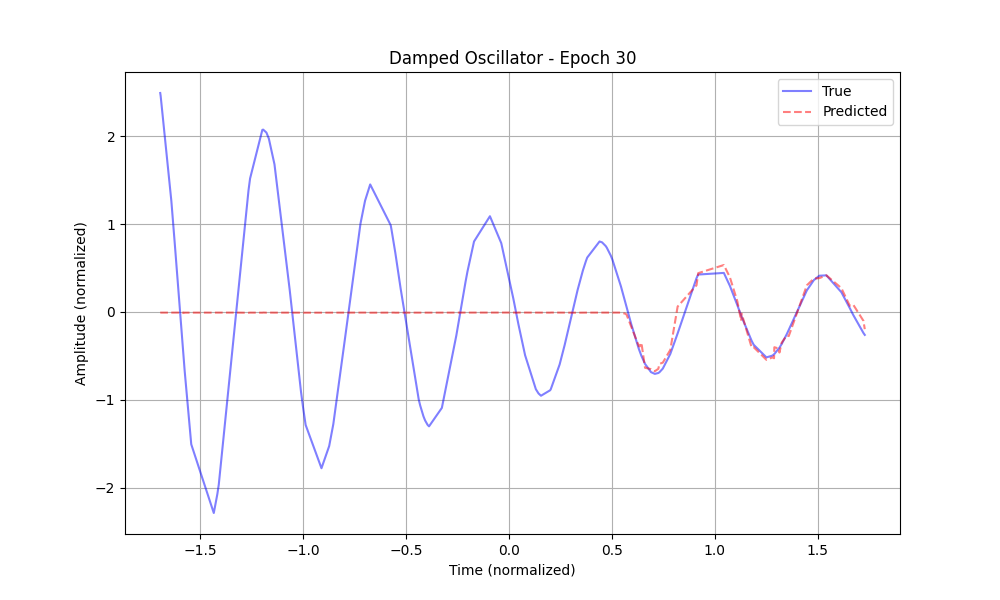}
    \caption{Traditional Transformer - Epoch 30}
  \end{subfigure}

  \begin{subfigure}[b]{0.3\textwidth}
    \includegraphics[width=\textwidth]{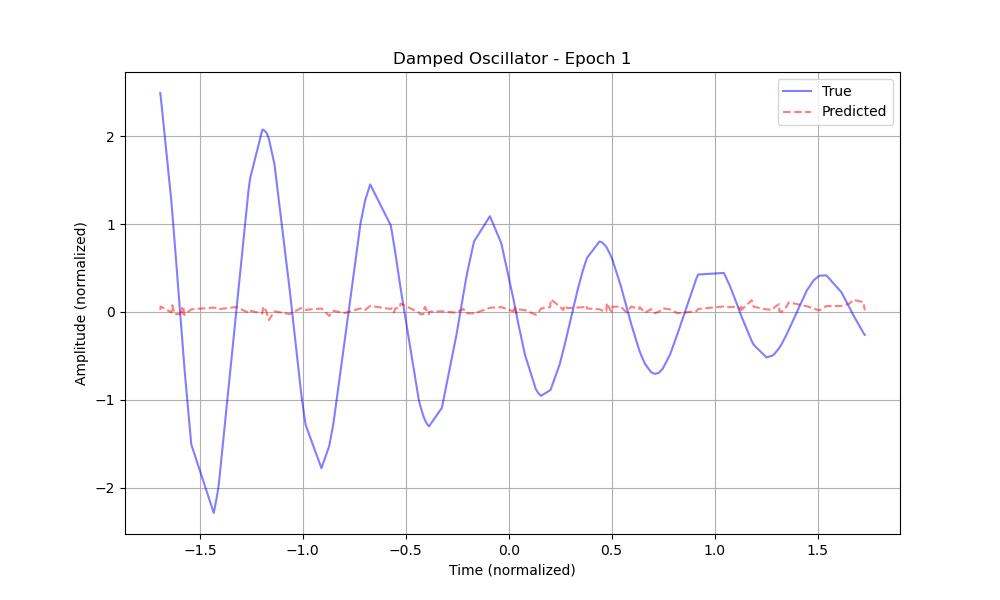}
    \caption{$QASA_{classical}$ - Epoch 1}
  \end{subfigure}
  \begin{subfigure}[b]{0.3\textwidth}
    \includegraphics[width=\textwidth]{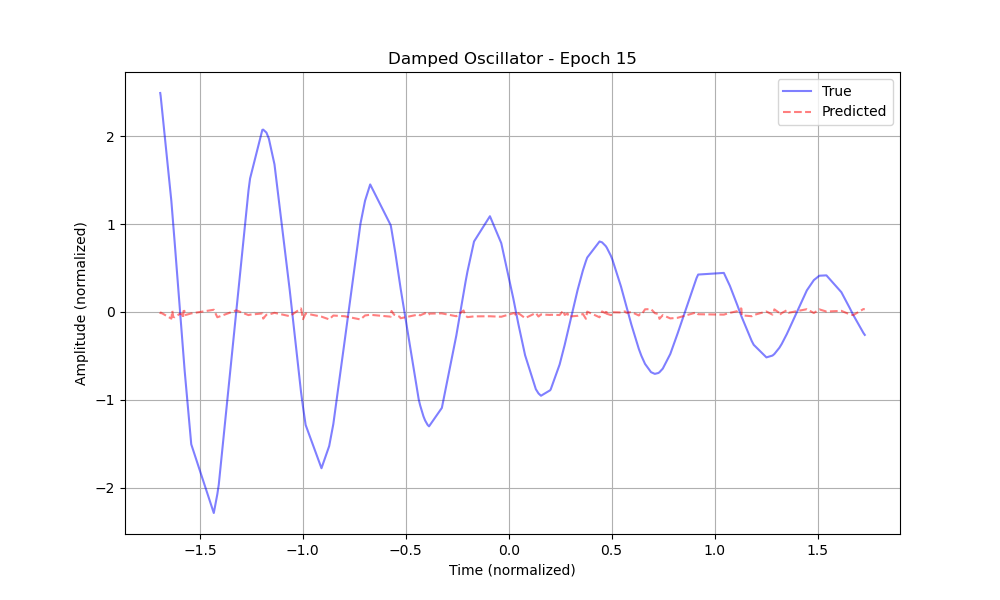}
    \caption{$QASA_{classical}$ - Epoch 15}
  \end{subfigure}
  \begin{subfigure}[b]{0.3\textwidth}
    \includegraphics[width=\textwidth]{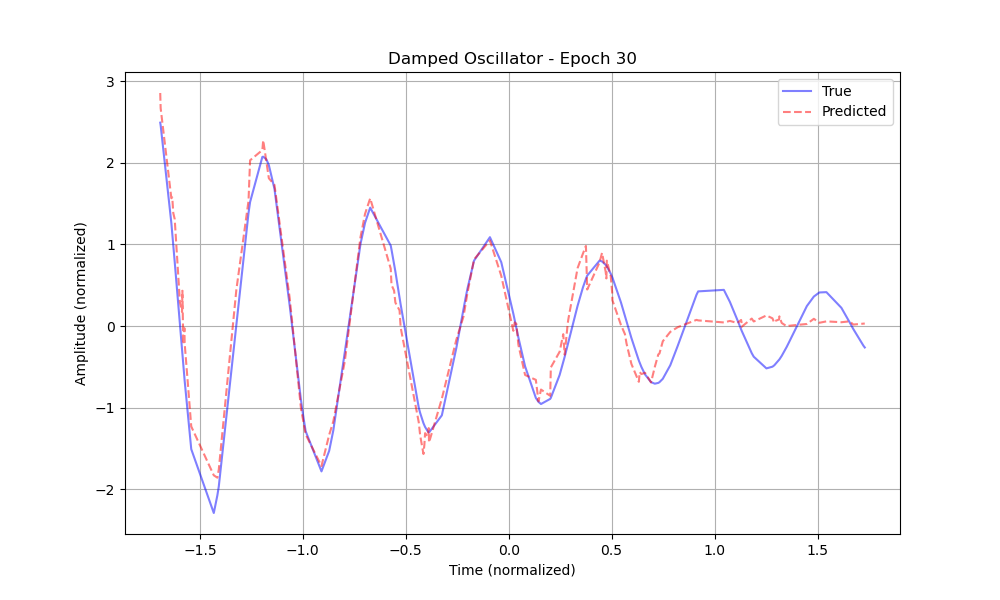}
    \caption{$QASA_{classical}$ - Epoch 30}
  \end{subfigure}

  \begin{subfigure}[b]{0.3\textwidth}
    \includegraphics[width=\textwidth]{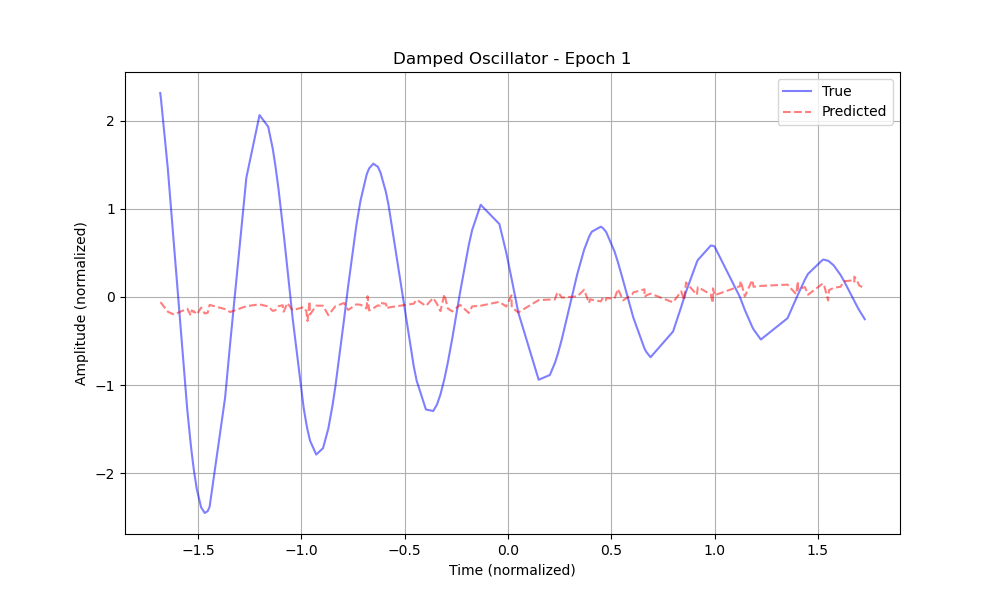}
    \caption{$QASA$ - Epoch 1}
  \end{subfigure}
  \begin{subfigure}[b]{0.3\textwidth}
    \includegraphics[width=\textwidth]{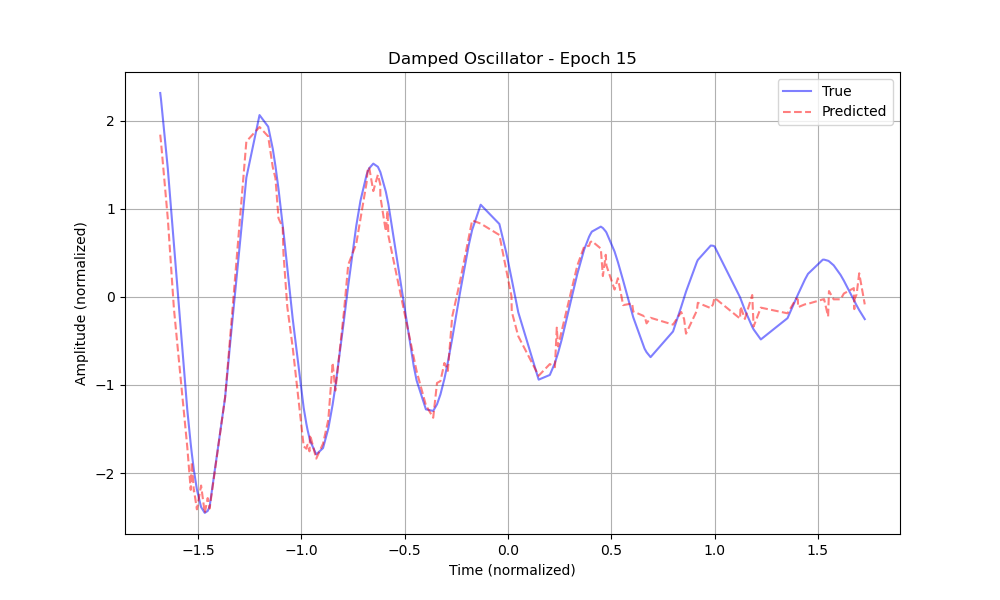}
    \caption{$QASA$ - Epoch 15}
  \end{subfigure}
  \begin{subfigure}[b]{0.3\textwidth}
    \includegraphics[width=\textwidth]{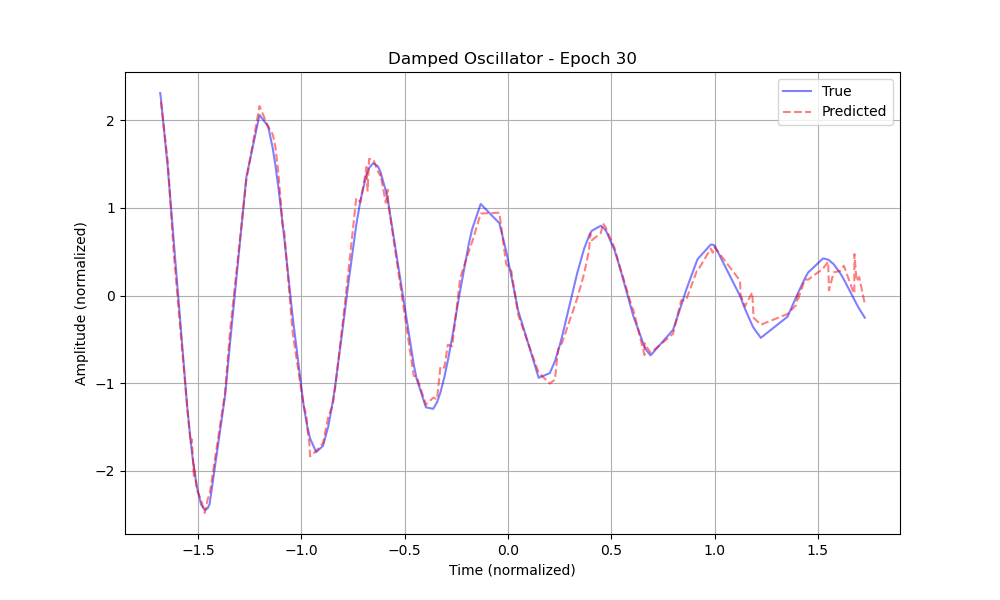}
    \caption{$QASA$ - Epoch 30}
  \end{subfigure}

  \caption{
Visual comparison of prediction performance across three models---Traditional Transformer, \( \text{QASA}_{\text{classical}} \), and QASA---at epochs 1, 15, and 30 on the damped oscillator task. At epoch 1, all models show poor predictions. By epoch 15, QASA already demonstrates significantly improved alignment with the true signal, while the other models lag behind. At epoch 30, QASA achieves near-perfect predictions, indicating much faster convergence compared to both the classical and transformer baselines.
}
\end{figure}

\paragraph{Gate Composition in QuantumLayer.}

The design of the parameterized quantum circuit (PQC) in the \texttt{QuantumLayer} follows a single-encoding, entanglement-aware structure optimized for hybrid neural architectures. Each layer of the PQC is composed of three primary stages:

\begin{enumerate}
    \item \textbf{Data Encoding via Single-Qubit Rotations:} The input features are encoded into quantum states using $R_X$ and $R_Z$ gates per qubit:
    \begin{equation}
    \forall i \in \{0, \dots, n-1\}, \quad \text{Apply } R_X(x_i),\ R_Z(x_i) \text{ on wire } i.
    \end{equation}

    This approach ensures both amplitude and phase information from classical features are embedded into the quantum state.

    \item \textbf{Parameterized Variational Rotations:} Across $L_q$ layers, the circuit includes learnable $R_Y$ and $R_Z$ rotations:
    \begin{equation}
    R_Y(\theta_{l, i})\ R_Z(\theta_{l, i}) \quad \forall i, \forall l \in \{1, \dots, L_q\},
    \end{equation}
    enabling expressive nonlinear transformations. These are \emph{trainable} rotations (parameters $\theta_{l,i}$); they are distinct from \emph{data} re-uploading---re-encoding the input $x$ before each layer---which we study as a separate encoding variant in Section~\ref{sec:encoding}. The default QASA circuit uses the single input encoding above, not data re-uploading.

    \item \textbf{Entanglement via Circular CNOT Topology:} To capture feature interactions across qubits, entanglement is introduced via a ring of CNOT gates:
    \begin{equation}
    \text{CNOT}(i \rightarrow (i+1) \bmod n),\quad \text{for } i = 0, \dots, n{-}1.
    \end{equation}

    An additional CNOT is applied from the final qubit to an auxiliary $(n+1)$-th qubit, allowing enhanced control or global interaction effects:
    \begin{equation}
    \text{CNOT}(n{-}1 \rightarrow n),\quad R_Y(\theta_{l, n}) \text{ on wire } n.
    \end{equation}
\end{enumerate}

This gate configuration balances expressiveness and hardware feasibility. The use of simple universal gates ($R_X, R_Y, R_Z$, CNOT) ensures compatibility with most current quantum hardware, while circular entanglement introduces full connectivity with only $n$ CNOTs per layer, avoiding unnecessary depth. The auxiliary $(n+1)$-th qubit provides an additional entangling degree of freedom coupled to the data register. (We make no claim that it captures \emph{long-range temporal} dependencies---those are handled by the classical self-attention; the ancilla acts only within the per-token quantum value map.)

\paragraph{Two circuits, two parameter-counting conventions.} To prevent confusion between tables, we distinguish up front the \emph{deployed} PQC (used in every experiment) from the \emph{depth-sweep reference} circuit used only for the barren-plateau trainability study (Table~\ref{tab:barren}). They are parameterised differently and their angle counts should never be added or directly compared (Table~\ref{tab:circuit_counts}).

\begin{table}[h]
\centering

\caption{The two circuit objects in this paper. The deployed PQC is fixed across all experiments; the depth-sweep reference exists only to probe trainability vs.\ depth (Table~\ref{tab:barren}). ``Variational layers'' counts trainable rotation blocks; one $R_X/R_Z$ encoding-rotation block plus three entangling blocks gives the deployed $4\times9=36$ angles and $3\times9=27$ CNOTs.}
\label{tab:circuit_counts}
\begin{tabular}{lcccc}
\toprule
\textbf{Circuit} & \textbf{Where used} & \textbf{Variational layers} & \textbf{Trainable angles} & \textbf{CNOTs} \\
\midrule
Deployed PQC & all experiments & 4 (1 rot.\ + 3 entangling) & $36$ & $27$ \\
Depth-sweep ref.\ & Table~\ref{tab:barren} only & 1 / 2 / 4 entangling & $18$ / $27$ / $45$ & varies \\
\bottomrule
\end{tabular}

\end{table}

\subsection{Circuit Expressibility and Entanglement Analysis}

To characterize the representational capacity of the QASA quantum circuit beyond its architectural description, we evaluate three established metrics from the quantum circuit benchmarking literature: entanglement entropy, the Meyer--Wallach entangling capability measure, and expressibility.

\paragraph{Entanglement Entropy.}
We compute the bipartite von Neumann entropy $S(\rho_A) = -\tr(\rho_A \log_2 \rho_A)$ across a near-balanced bipartition of the qubits (qubits 0--3 vs.\ 4--7 plus auxiliary, i.e.\ 4 vs.\ 5 qubits), averaged over 200 random parameter samples. The QASA circuit achieves a mean entanglement entropy of $2.26 \pm 0.12$ bits out of a maximum of 4.0 bits (normalized: 0.57), indicating substantial but not maximal entanglement. This moderate level is desirable: it ensures non-trivial quantum correlations across qubit subsets without collapsing into a maximally entangled state that would limit trainability~\cite{mcclean2018barren}.

\paragraph{Meyer--Wallach Entangling Capability.}
The Meyer--Wallach measure~\cite{meyer2002global} quantifies the global entanglement of a pure state $|\psi\rangle$ on $N$ qubits as
\begin{equation}
\label{eq:mw}
Q(|\psi\rangle) = 2 - \frac{2}{N}\sum_{k=1}^{N} \tr\!\left(\rho_k^2\right),
\end{equation}
where $\rho_k = \tr_{\neg k}\!\left(|\psi\rangle\langle\psi|\right)$ is the reduced density matrix of qubit $k$. By construction, $Q \in [0, 1]$: $Q = 0$ for product states and $Q = 1$ when every single-qubit reduced density matrix is maximally mixed. Averaged over 1{,}000 random parameterizations, QASA achieves $Q = 0.981$, substantially higher than QLSTM ($Q = 0.710$) and QnnFormer ($Q = 0.883$). This confirms that the ring-topology CNOT structure with auxiliary qubit coupling generates near-maximally entangled states, despite using fewer two-qubit gates than QLSTM (27 vs.\ 56 CNOTs). The high $Q$ and the moderate bipartite entropy above are consistent rather than contradictory: $Q$ averages single-qubit mixedness over all qubits (sensitive to widely distributed entanglement), whereas the entropy measures correlation across one $4$-vs-$5$ cut; together they indicate strong, broadly distributed entanglement that is nonetheless sub-maximal across any single balanced partition.

\paragraph{Expressibility.}
Following Sim et al.~\cite{sim2019expressibility}, we quantify the expressibility of a PQC $U(\theta)$ as the Kullback--Leibler divergence between the distribution of pairwise state fidelities generated by random circuit parameters and the Haar-random fidelity distribution:
\begin{equation}
\label{eq:expressibility}
D_{\mathrm{KL}} = D_{\mathrm{KL}}\!\left(\,P_{\mathrm{PQC}}(F) \,\|\, P_{\mathrm{Haar}}(F)\,\right),
\qquad F = \left|\langle \psi(\theta_1) \,|\, \psi(\theta_2)\rangle\right|^2,
\end{equation}
where the Haar reference on $N$ qubits is $P_{\mathrm{Haar}}(F) = (d - 1)(1 - F)^{d-2}$ with $d = 2^N$. We sample 5{,}000 random parameter pairs $(\theta_1, \theta_2)$ and estimate $P_{\mathrm{PQC}}$ via a 75-bin histogram. QASA yields $D_{\mathrm{KL}} = 0.029$, comparable to QnnFormer ($D_{\mathrm{KL}} = 0.026$) and substantially better than QLSTM ($D_{\mathrm{KL}} = 0.125$). Lower $D_{\mathrm{KL}}$ indicates wider coverage of the Hilbert space, a necessary condition for effective variational learning.

\begin{table}[h]
\centering
\caption{Comparative quantum circuit analysis across all three quantum models (5{,}000 parameter samples for expressibility, 1{,}000 for entangling capability). Lower expressibility KL divergence indicates wider Hilbert space coverage; higher Meyer--Wallach $Q$ indicates stronger entanglement generation. Counting convention for the deployed QASA circuit (9 wires = 8 data + 1 ancilla): the trainable weight tensor is $4\times9=36$ angles (one $R_X/R_Z$ rotation layer followed by 3 entangling layers); the 27 CNOTs are the 3 entangling layers $\times$ (8 ring + 1 ancilla) gates. The differently-parameterised depth-sweep circuit in Table~\ref{tab:barren} is a separate trainability reference and should not be conflated with these 36 angles. The \textbf{Q-Params} column reports the parameters of the single representative circuit analysed here; at the \emph{model} level the quantum-parameter budgets (Table~\ref{tab:baseline_mae}) are $36$ (QASA), $128$ (QLSTM, which instantiates one such circuit per LSTM gate), and $90$ (QnnFormer), so QASA uses the \emph{fewest} quantum parameters of the three at the model level, despite the larger per-circuit count shown in this row.}
\label{tab:circuit_analysis}
\resizebox{\columnwidth}{!}{
\begin{tabular}{lcccccc}
\toprule
\textbf{Circuit} & \textbf{Qubits} & \textbf{Q-Params} & \textbf{CNOT Gates} & \textbf{Total Gates} & \textbf{Expr.\ ($D_{\mathrm{KL}}\downarrow$)} & \textbf{Ent.\ Cap.\ ($Q\uparrow$)} \\
\midrule
QASA & 9 & 36 & 27 & 110 & 0.029 & \textbf{0.981} \\
QLSTM & 8 & 32 & 56 & 104 & 0.125 & 0.710 \\
QnnFormer & 8 & 24 & 21 & 94 & \textbf{0.026} & 0.883 \\
\bottomrule
\end{tabular}
}
\end{table}

Table~\ref{tab:circuit_analysis} reveals a striking efficiency gap: QASA achieves the highest entangling capability ($Q = 0.981$) with only 27 CNOT gates, while QLSTM requires more than twice as many CNOTs (56) yet produces the weakest entanglement ($Q = 0.710$). This demonstrates that QASA's ring topology---where each qubit is entangled with its neighbor in a circular arrangement---is fundamentally more efficient than QLSTM's linear chain topology. QnnFormer achieves competitive expressibility with the fewest total gates (94), but its entangling power remains below QASA's. These results provide a circuit-level explanation for QASA's strong empirical performance: near-maximal entanglement enables the quantum layer to capture complex correlations, while high expressibility ensures the circuit can represent diverse functions during training.

\subsection{Prediction and Training Objective}

The prediction is generated by extracting the final time step representation $h_N[L] \in \mathbb{R}^d$ and applying a linear projection:
\begin{equation}
\hat{y} = W_{out} \cdot h_N[L] + b_{out}.
\end{equation}

The model is trained using the Mean Squared Error (MSE) loss:
\begin{equation}
\mathcal{L}_{\text{MSE}} = \frac{1}{N} \sum_{i=1}^N (y_i - \hat{y}_i)^2.
\end{equation}

Optimization is performed with AdamW and a cosine annealing learning rate scheduler. We apply early stopping and save the model with the best validation loss. During training, we also visualize sorted and unsorted predictions to monitor performance.

\section{Results and Discussions}

\begin{table}[h]
\centering
\caption{Comparison of model architectures: $QASA_{classical}$ and $QASA$.}
\label{tab:qasa-compare}
\resizebox{\columnwidth}{!}{
\begin{tabular}{ l c c }
\hline
\textbf{Component} & \textbf{$QASA_{classical}$} & \textbf{$QASA$} \\
\hline
Input Projection & Linear($1 \rightarrow 256$) & Linear($1 \rightarrow 256$) + LayerNorm \\
Positional Encoding & Sinusoidal & Sinusoidal \\
Encoder Layers & 4$\times$ Transformer (4 heads) & 3$\times$ Transformer + 1$\times$ QuantumEncoderLayer \\
Feedforward Dim & 1024 & 1024 (post-quantum FFN) \\
Activation Function & GELU & GELU \\
Quantum Component & \xmark & \cmark\ 8-qubit, 4-layer VQC \\
Output Layer & 2-layer MLP (GELU + Linear) & Linear(256 $\rightarrow$ 1) \\
Attention Heads & 4 & 4 \\
Hidden Dim & 256 & 256 \\
\hline
\end{tabular}
}
\end{table}

\subsection{Experiment Details}
\begin{figure}
    \centering
    \includegraphics[width=0.9\textwidth]{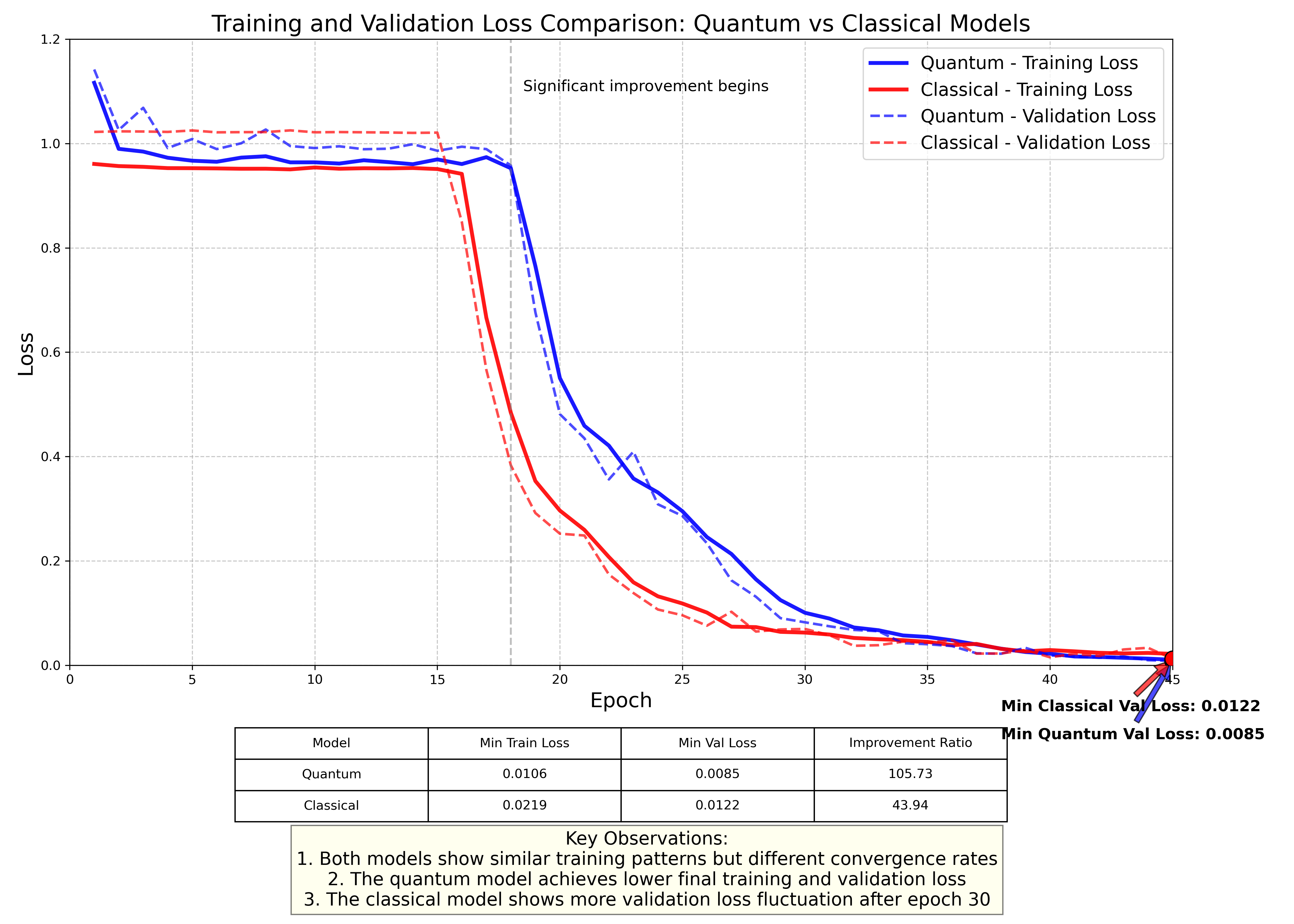}
    \caption{
\emph{Preliminary, single-run motivating observation} (larger hidden dimension $256$, 45 epochs). Comparison of training and validation loss between the classical model \( \text{QASA}_{\text{classical}} \) and the quantum model QASA.  The quantum model reaches a lower minimum validation loss ($0.0085$) than the classical model ($0.0122$) with smoother convergence.  This single-run observation is what originally motivated the present study; we stress it is \emph{not} evidence of a quantum advantage. The systematic, multi-seed, capacity-matched analysis below (Table~\ref{tab:bottleneck}) shows that such gains over a full-capacity classical model are attributable to the low-rank value bottleneck rather than to quantumness. We retain this figure only as the motivating observation.}

    \label{fig:loss_comparison}
\end{figure}

To evaluate the effectiveness of quantum neural network (QNN) integration in time-series regression, we compare three transformer-based architectures under identical training settings:

\begin{itemize}
    \item \textbf{Transformer}: A standard transformer architecture consisting of an input projection layer, sinusoidal positional encoding, four unmodified transformer encoder blocks (each with 8 attention heads and a feed-forward network of dimension 1024), and an output MLP head. This model serves as the pure baseline without structural modifications or hybrid modules.

    \item \textbf{$QASA_{classical}$}: A classical transformer-based variant similar to the vanilla model but with reduced architectural complexity. It includes 4 transformer encoder blocks, each with 4 attention heads and a hidden size of 256. The output head uses GELU activation and layer normalization to align with the structure of the quantum model.

    \item \textbf{$QASA$}: A quantum-enhanced hybrid transformer in which the final transformer encoder block is replaced with a quantum encoder layer that incorporates a variational quantum circuit (VQC). The VQC uses 8 qubits and 4 layers, with $R_X$ and $R_Z$ rotations and entangling CNOT gates, and outputs expectation values of Pauli-$Z$ measurements. These values are projected back into the hidden dimension space and processed in a residual feed-forward fashion.
\end{itemize}

All models are trained on a synthetic damped oscillator prediction task, defined by:
\begin{equation}
    x(t) = A e^{-\gamma t} \cos(\omega t + \phi)
\end{equation}

with parameters $A = 1.0$, $\gamma = 0.1$, $\omega = 2.0$, and $\phi = 0$. The models are trained to predict the next amplitude given a sequence of 50 normalized time steps. The data is standardized, and an 80/20 train-validation split is used.

Training is conducted using PyTorch Lightning with the AdamW optimizer (learning rate $1 \times 10^{-4}$, weight decay $0.01$), and a ReduceLROnPlateau or cosine annealing scheduler. Each model is trained for 45 epochs, with early stopping based on validation loss. Metrics including mean squared error (MSE), mean absolute error (MAE), and total loss are logged during training. Visualization of predicted vs.\ true amplitudes at selected epochs provides additional insight into model convergence and performance.

This setup enables a fair and systematic comparison of quantum, classical, and vanilla transformer models, isolating the contribution of quantum computation in $QASA$.

For the broader benchmark evaluation (Table~\ref{tab:ts_mae_mse}), we use a reduced hidden dimension of 64 (with 4 attention heads and 4 encoder layers) to ensure tractable quantum circuit simulation across all nine tasks. Training is conducted for 200 epochs using the AdamW optimizer (learning rate $5 \times 10^{-4}$, weight decay $10^{-4}$) with cosine annealing scheduling. Each configuration is evaluated over five random seeds to assess consistency. Under this configuration, the QASA model has 201,405 total parameters and the classical Transformer baseline has 200,257 parameters---a difference of less than 0.6\%, ensuring a fair parameter-matched comparison.

\textbf{A note on the classical baselines.} To avoid confusion, three distinct classical references appear in this paper, each with a specific role: (i) the \emph{full-capacity Classical Transformer} (hidden\_dim$=64$, $200{,}257$ parameters)---the primary parameter-matched baseline used throughout the nine-task benchmark (Tables~\ref{tab:ts_mae_mse} and~\ref{tab:baseline_mae}); (ii) the \emph{classical bottleneck} (Table~\ref{tab:bottleneck})---identical to QASA but with the PQC replaced by a $40$-parameter classical low-rank value map, used to isolate the quantum contribution from the bottleneck structure; and (iii) $QASA_{classical}$ (hidden\_dim$=256$)---a reduced variant used \emph{only} in the preliminary motivating comparison of Figure~\ref{fig:loss_comparison}. Unless stated otherwise, ``Classical'' in the results refers to~(i).

\paragraph{Experimental protocol and why some numbers differ across tables.} For transparency we collect here the protocol choices that explain apparent number differences between tables, so they are not mistaken for inconsistencies. (i) \emph{Evaluation mode.} All error-metric tables (Tables~\ref{tab:ts_mae_mse},~\ref{tab:bottleneck},~\ref{tab:baseline_mae},~\ref{tab:baseline_mse}) report \emph{free-running autoregressive rollout} over the test horizon (each prediction is fed back as input), which compounds error---especially on chaotic signals---and is why absolute MAEs (${\sim}0.3$--$0.4$ on chaotic) are far larger than the \emph{one-step teacher-forced} MAEs of the hardware run (Section~\ref{sec:hardware}, ${\sim}0.01$) and the teacher-forced train/test ratios in the overfitting analysis. (ii) \emph{Seed count.} The main benchmark uses 5 seeds (42--46); the quantum-baseline and representation analyses use 3; the encoding/sequence-length/qubit sweeps use 1--2. Small-error clean-periodic tasks have high relative seed variance, so e.g.\ the damped-oscillator MAE differs between the 5-seed Table~\ref{tab:ts_mae_mse} ($0.117$) and the 3-seed Table~\ref{tab:baseline_mae} ($0.086$); both use the identical hidden\_dim$=64$/200-epoch setup. (iii) \emph{Reduced circuits and budgets.} The noise (Tables~\ref{tab:noise},~\ref{tab:noise_multi}) and some sweep experiments use a smaller PQC (4-qubit/2-layer) and/or fewer epochs for tractability, so their noiseless baselines (e.g.\ $0.384$ vs.\ $0.333$ across the two noise tables, at 3 vs.\ 2 seeds) are not comparable across tables, only within. Each table's caption states its own protocol; comparisons are valid \emph{within} a table.

\paragraph{Computational Cost.}
All experiments are conducted on a single CPU (Apple M-series) using PennyLane's \texttt{lightning.qubit} simulator for the quantum circuit components. Table~\ref{tab:training_time} reports the wall-clock training time per task (200 epochs) averaged across the nine benchmark tasks. The classical Transformer completes training in approximately 4.5 minutes per task (1.4\,s/epoch), while QASA requires approximately 2 hours and 7 minutes per task (40\,s/epoch)---roughly a $\mathbf{29\times}$ slowdown. This overhead is entirely attributable to quantum circuit \textit{simulation}: each forward pass through the QuantumLayer requires executing the parameterized quantum circuit for every token in the sequence, with gradient computation via the parameter-shift rule doubling the number of circuit evaluations. On actual quantum hardware, circuit execution time scales with circuit depth ($L_q \times$ gate time) rather than with the exponential cost of classical state-vector simulation, and the overhead would be substantially reduced.

\begin{table}[h]
\centering
\caption{Average wall-clock training time per task (200 epochs, single CPU, hidden\_dim=64). The $29\times$ overhead arises from simulating the 9-qubit PQC classically; on quantum hardware, circuit execution would be substantially faster.}
\label{tab:training_time}
\begin{tabular}{l c c}
\hline
\textbf{Model} & \textbf{Time/task} & \textbf{sec/epoch} \\
\hline
Classical Transformer & 4.5 min & 1.4\,s \\
QASA  & 2h\,07m & 40.4\,s \\
\hline
Slowdown factor & \multicolumn{2}{c}{$\approx 29\times$} \\
\hline
\end{tabular}
\end{table}

\subsection{Dataset Construction}
\label{sec:datasets}

To probe the task-conditional behaviour of quantum-enhanced attention, we construct nine univariate synthetic time-series benchmarks, each designed to expose a distinct signal regime---smooth periodic, chaotic, stochastic, trend-dominated, discontinuous, and mixed. All sequences contain $T = 500$ points and are split chronologically into 80\% training and 20\% testing. For all tasks, the model is trained to predict $x_{t+1}$ from a sliding context window of length $L = 20$. Real-world validation is performed on the ETTh1 dataset; see the corresponding paragraph below.

\paragraph{Synthetic tasks.} Let $t \in \{0, 1, \dots, T{-}1\}$ denote the discrete time index and $\tau = 50\,t/(T{-}1)$ the normalised continuous time on $[0, 50]$. The nine signals are defined as follows:

\begin{enumerate}[leftmargin=*,itemsep=2pt]
    \item \textbf{ARMA}: A second-order autoregressive moving-average process,
    \begin{equation}
        x_t = 0.75\,x_{t-1} - 0.25\,x_{t-2} + \varepsilon_t + 0.65\,\varepsilon_{t-1}, \qquad \varepsilon_t \sim \mathcal{N}(0, 1).
    \end{equation}

    \item \textbf{Chaotic logistic map}: A deterministic chaotic system,
    \begin{equation}
        x_t = r\, x_{t-1}(1 - x_{t-1}), \qquad r = 3.9,\ x_0 = 0.2.
    \end{equation}
    The parameter $r = 3.9$ places the system deep in the chaotic regime.

    \item \textbf{Damped oscillator}: A clean decaying sinusoid,
    \begin{equation}
        x(\tau) = e^{-0.05\,\tau}\sin(\tau).
    \end{equation}
    (This is the benchmark generator used for the nine-task suite at $L=20$; it differs from the preliminary damped signal $x(t)=e^{-0.1t}\cos(2t)$ of the Experiment Details subsection, which is used \emph{only} for the motivating comparison in Figure~\ref{fig:loss_comparison} and the epoch snapshots.)

    \item \textbf{Noisy damped oscillator}: As above with additive Gaussian noise,
    \begin{equation}
        x_t = e^{-0.05\,t}\sin(2\pi \cdot 0.2\,t) + \eta_t, \qquad \eta_t \sim \mathcal{N}(0, 0.05^2).
    \end{equation}

    \item \textbf{Piecewise regime}: A concatenation of six linear regimes with two impulsive shocks at $t \in [200, 205)$ and $t \in [350, 355)$, mimicking structural breaks,
    \begin{equation}
        x_t = \alpha_k\,t + \beta_k + \eta_t, \quad \eta_t \sim \mathcal{N}(0, \sigma_k^2), \ (\alpha_k, \beta_k, \sigma_k) \text{ regime-dependent}.
    \end{equation}
    The exact per-regime $(\alpha_k,\beta_k,\sigma_k)$ are listed in Appendix~\ref{app:generators}, Table~\ref{tab:piecewise} (segments 4--5 use $\sigma_k=0.02$, the others $0.01$).

    \item \textbf{Sawtooth wave}: A periodic ramp with period $P = 10$,
    \begin{equation}
        x(\tau) = 2\left(\frac{\tau}{P} - \left\lfloor \tfrac{1}{2} + \frac{\tau}{P} \right\rfloor\right).
    \end{equation}

    \item \textbf{Square wave}: A binary waveform with time-varying period and duty cycle,
    \begin{equation}
        x(\tau) = \mathrm{sgn}\!\left( d(\tau) - \phi(\tau) \right),
    \end{equation}
    where $\phi(\tau) = (\tau \bmod P(\tau))/P(\tau)$, $P(\tau) = 5 + 5\sin(0.1\tau)$, and $d(\tau) = 0.2 + 0.3\left(1 + \sin(0.05\tau)\right)$.

    \item \textbf{Seasonal trend with noise}: An additive mixture of trend, seasonality, and noise,
    \begin{equation}
        x_t = 0.03\,t + \sin(0.4\,t) + \eta_t, \qquad \eta_t \sim \mathcal{N}(0, 0.2^2).
    \end{equation}

    \item \textbf{Waveform}: A smooth reference sinusoid, $x(\tau) = \sin(\tau)$.
\end{enumerate}

These nine tasks span four qualitatively different dynamical regimes: (i) \emph{stochastic but stationary} (ARMA), (ii) \emph{chaotic} (logistic map), (iii) \emph{smooth periodic/quasi-periodic} (damped oscillator, waveform, sawtooth, square wave), and (iv) \emph{non-stationary or discontinuous} (piecewise regime, seasonal trend, noisy damped oscillator). This coverage lets us identify \emph{when} the low-rank bottleneck helps rather than making a single aggregate claim.

\paragraph{Real-world dataset (ETTh1).} The Electricity Transformer Temperature dataset~\cite{zhou2021informer} provides hourly measurements of oil temperature from an electricity transformer over approximately two years. We use a subsampled training set of 500 windows (sequence length $L = 20$) with the target variable ``OT'' (oil temperature), and evaluate on a held-out test set of 500 windows. No dataset-specific hyperparameter tuning is performed; all models use the same configuration as the synthetic benchmarks.

\subsection{Results}
Initial experiments with a larger model capacity (hidden\_dim=256, 45 epochs) suggested that QASA could outperform classical models on specific tasks, motivating the systematic parameter-matched comparison in Table~\ref{tab:ts_mae_mse}.

\begin{table}[h]
\centering
\caption{Performance comparison across nine benchmark tasks (5 seeds, mean$\pm$std). Both models share nearly identical parameter counts (Classical: 200,257; QASA: 201,405). Bold indicates the better mean result per metric per task.}
\label{tab:ts_mae_mse}
\resizebox{\columnwidth}{!}{
\begin{tabular}{lccccc}
\toprule
\textbf{Task} & \textbf{Model} & \textbf{MAE} & \textbf{MAE Std} & \textbf{MSE} & \textbf{MSE Std} \\
\midrule
ARMA & Classical & 2.1522 & 0.1410 & \textbf{7.0728} & 1.0097 \\
     & QASA     & \textbf{2.1517} & 0.3780 & 7.4180 & 2.5988 \\
\midrule
Chaotic Logistic & Classical & 0.3729 & 0.0373 & 0.2044 & 0.0256 \\
                 & QASA     & \textbf{0.3406} & 0.0519 & \textbf{0.1824} & 0.0345 \\
\midrule
Damped Oscillator & Classical & \textbf{0.0375} & 0.0204 & \textbf{0.0021} & 0.0018 \\
                  & QASA     & 0.1170 & 0.0810 & 0.0288 & 0.0344 \\
\midrule
Noisy Damped Osc & Classical & \textbf{0.0434} & 0.0010 & \textbf{0.0029} & 0.0001 \\
                 & QASA     & 0.0434 & 0.0016 & 0.0029 & 0.0002 \\
\midrule
Piecewise Regime & Classical & \textbf{22.1308} & 0.0965 & \textbf{490.5139} & 4.2751 \\
                 & QASA     & 22.5857 & 0.4630 & 511.0187 & 21.1518 \\
\midrule
Sawtooth & Classical & \textbf{0.0938} & 0.0418 & \textbf{0.0598} & 0.0652 \\
         & QASA     & 0.2268 & 0.1861 & 0.1961 & 0.2714 \\
\midrule
Square Wave & Classical & 0.8685 & 0.0899 & \textbf{0.9866} & 0.1830 \\
            & QASA     & \textbf{0.7946} & 0.1144 & 1.1185 & 0.2037 \\
\midrule
Seasonal Trend & Classical & 0.6739 & 0.0490 & 0.5705 & 0.0647 \\
               & QASA     & \textbf{0.6676} & 0.0104 & \textbf{0.5542} & 0.0122 \\
\midrule
Waveform & Classical & \textbf{0.0675} & 0.0313 & \textbf{0.0065} & 0.0053 \\
         & QASA     & 0.0894 & 0.0275 & 0.0110 & 0.0062 \\
\bottomrule
\end{tabular}
}
\end{table}

To evaluate the effectiveness of the proposed Quantum Adaptive Self-Attention (QASA) architecture, we conducted experiments across nine synthetic time-series forecasting tasks, each characterized by different signal dynamics---ranging from periodicity and chaos to noise and discontinuities. Both models share nearly identical parameter counts (Classical: 200,257; QASA: 201,405), ensuring a fair comparison. Each experiment was repeated over five random seeds and we report mean and standard deviation. Performance was assessed using Mean Absolute Error (MAE) and Mean Squared Error (MSE), as shown in Table~\ref{tab:ts_mae_mse}.

In the chaotic logistic map---a highly nonlinear and sensitive system---QASA outperformed its classical counterpart (MAE: 0.3406 vs.\ 0.3729; MSE: 0.1824 vs.\ 0.2044), with a small but consistent effect size (Cohen's $d = 0.40$); we note, however, that the parameter-matched bottleneck analysis (Table~\ref{tab:bottleneck}) attributes this gain over the full-capacity baseline to the low-rank value bottleneck rather than to quantumness per se. In the seasonal trend task, QASA achieved lower error with remarkably consistent performance across seeds (MAE: $0.6676{\pm}0.0104$ vs.\ $0.6739{\pm}0.0490$; MSE: $0.5542{\pm}0.0122$ vs.\ $0.5705{\pm}0.0647$), suggesting that quantum circuits provide stable representations for smooth trend-based patterns. In the square wave task, QASA achieved a lower MAE (0.7946 vs.\ 0.8685) with a medium effect size ($d = 0.56$), though its MSE was higher (1.1185 vs.\ 0.9866), suggesting that while the quantum model captures the general waveform shape more accurately on average, it may produce occasional large deviations at sharp transitions. In the ARMA task, QASA and the classical model achieved nearly identical MAE (2.1517 vs.\ 2.1522; $d = 0.002$), indicating negligible difference for structured autoregressive dynamics.

However, the performance gains did not generalize uniformly across all tasks. In the damped oscillator task, the classical model significantly outperformed QASA (MAE: 0.0375 vs.\ 0.1170; $d = -1.18$, $p = 0.058$), the largest effect size observed in our benchmark, confirming that for clean, low-frequency periodic signals the overhead of quantum encoding introduces unnecessary complexity. Similarly, in the piecewise regime task, the classical model showed a large advantage ($d = -1.01$), indicating that abrupt temporal discontinuities remain challenging for quantum-enhanced attention. The sawtooth and waveform tasks also favored the classical model with large and medium effect sizes (Sawtooth $d = -0.99$; Waveform $d = -0.52$). The noisy damped oscillator task showed essentially identical performance between both models (MAE: 0.0434 vs.\ 0.0434; $d = -0.01$).

Overall, by MAE, QASA outperforms the classical baseline in four out of nine tasks (ARMA, Chaotic Logistic, Square Wave, and Seasonal Trend), while the classical model prevails in the remaining five. By MSE, QASA wins two tasks (Chaotic Logistic and Seasonal Trend). Paired $t$-tests (5 seeds) did not reach statistical significance at $\alpha = 0.05$ for any individual task, with the closest being the damped oscillator ($p = 0.058$); however, Cohen's $d$ effect sizes reveal meaningful patterns: the classical model achieves large advantages on clean periodic and discontinuous signals ($|d| > 0.8$), while QASA shows small-to-medium advantages on chaotic and composite signals ($d \approx 0.4$--$0.6$). QASA performs best on tasks characterized by nonlinear dynamics, chaotic behavior, or composite trend-seasonal patterns, where a low-rank value bottleneck---realised here quantumly---provides a representational advantage over the full-capacity classical baseline (the advantage is not specific to quantumness; see Table~\ref{tab:bottleneck}). Conversely, classical attention mechanisms remain more effective for clean periodic signals, structured autoregressive patterns (by MSE), and sharply discontinuous waveforms. These findings highlight both the promise and the task-specific nature of quantum-enhanced attention, suggesting that future work should focus on adaptive hybrid strategies that selectively engage quantum layers based on signal characteristics.

\paragraph{Is the gain quantum, or just a compact bottleneck?} A critical confound must be addressed: QASA's quantum value map is a low-capacity ($36$-parameter) bottleneck, and the gains over the full-capacity classical baseline could stem from this \emph{compactness} rather than from quantum feature mapping. To isolate the two, we construct a \textbf{parameter-matched classical bottleneck}: the architecture is identical to QASA in every respect---same three classical Transformer layers, attention, FFN, residual structure, optimiser, schedule, and seeds---except that the PQC value map is replaced by a classical low-rank map (rank-2 bilinear layer with a bounded $\tanh$ nonlinearity, $40$ trainable parameters vs.\ the PQC's $36$; total model $201{,}409$ vs.\ QASA's $201{,}405$). The \emph{only} difference is quantum vs.\ classical at the value bottleneck. Table~\ref{tab:bottleneck} reports the comparison (5 seeds).

\begin{table}[h]
\centering

\caption{Parameter-matched bottleneck analysis (5 seeds, matching Table~\ref{tab:ts_mae_mse}). The classical bottleneck replaces QASA's $36$-parameter PQC with a $40$-parameter classical low-rank value map, holding everything else fixed. The bottleneck matches or improves on the full classical model for the three nonlinear/composite tasks and is worse on the clean-periodic control; QASA is statistically \emph{indistinguishable} from the classical bottleneck on the three favourable tasks and clearly worse on the control. This isolates the gain as coming from the low-rank \emph{bottleneck structure}, not from quantumness. We reproduced the qualitative pattern on a second, independent 5-seed set (seeds $47$--$51$).}
\label{tab:bottleneck}
\begin{tabular}{lccc}
\toprule
\textbf{Task (MAE / MSE)} & \textbf{Classical (200k)} & \textbf{Classical bottleneck (40)} & \textbf{QASA (36 q)} \\
\midrule
Chaotic Logistic  & $0.373 / 0.204$ & $0.344 / 0.186$ & $\mathbf{0.341 / 0.182}$ \\
Seasonal Trend    & $0.674 / 0.571$ & $\mathbf{0.666 / 0.553}$ & $0.668 / 0.554$ \\
Square Wave       & $0.869 / \mathbf{0.987}$ & $\mathbf{0.802} / 1.127$ & $0.795 / 1.119$ \\
Damped Osc.\ (control) & $\mathbf{0.038 / 0.002}$ & $0.064 / 0.009$ & $0.117 / 0.029$ \\
\bottomrule
\end{tabular}

\end{table}

The conclusion is clear. On the three tasks where QASA improves on the full-capacity classical baseline (chaotic logistic, seasonal trend, square wave), the \emph{classical} bottleneck matches QASA to within seed variance, and on the clean-periodic control it is markedly better ($0.064$ vs.\ $0.117$ MAE) because the quantum map injects estimation noise where no compression is needed. The differences among the three models on the favourable tasks are small relative to the seed standard deviation, so we do \emph{not} claim that the quantum layer provides a task-level MAE/MSE advantage over a capacity-matched classical bottleneck; we reproduced this qualitative pattern on a second, independent 5-seed set (seeds $47$--$51$). Instead, the gain over the standard baseline is attributable to \textbf{architectural parsimony itself}---a single low-rank value bottleneck at the final encoder position---which both classical and quantum realisations exploit. This reframing strengthens rather than weakens the contribution: the parsimony principle is the robust finding, and (as we show next) the quantum realisation remains a \emph{competitive} instantiation distinguished by its physical properties---near-maximal entanglement with few gates, NISQ noise resilience verified on real hardware (Section~\ref{sec:hardware}), and barren-plateau-aware trainability---rather than by a distinct compression mechanism, which a classical bottleneck matches (Section~\ref{sec:datasets} representation analysis).

\paragraph{Real-World Validation on ETTh1.}
To assess whether the task-specific advantages observed on synthetic data extend to real-world settings, we evaluate both models on the ETTh1 (Electricity Transformer Temperature) dataset~\cite{zhou2021informer}, a widely used benchmark for time-series forecasting. ETTh1 contains hourly measurements of oil temperature from an electricity transformer over approximately two years. We use a subsampled training set of 500 windows (sequence length 20) with the target variable ``OT'' (oil temperature), and evaluate on a held-out test set of 500 windows. The same model configuration (hidden\_dim=64, 4 layers, 200 epochs) is used as in the synthetic benchmarks.

\begin{table}[h]
\centering
\caption{Performance on the real-world ETTh1 dataset (oil temperature forecasting). All models use identical configurations (hidden\_dim=64, 200 epochs, seq.\ len.\ 20, seed 42). Bold indicates the best result per metric. The capacity-matched classical bottleneck is the best on both metrics, confirming that the real-world gain---like the synthetic one---comes from the low-rank value bottleneck rather than quantumness.}
\label{tab:etth1}
\begin{tabular}{l c c}
\hline
\textbf{Model} & \textbf{MAE} & \textbf{MSE} \\
\hline
Classical Transformer (full, 200k) & 0.0718 & 0.0078 \\
Classical bottleneck (40 params) & $\mathbf{0.0621}$ & $\mathbf{0.0068}$ \\
QASA (36 q-params) & 0.0675 & 0.0079 \\
\hline
\end{tabular}
\end{table}

As shown in Table~\ref{tab:etth1}, QASA achieves a 6.0\% lower MAE (0.0675 vs.\ 0.0718) than the full-capacity classical Transformer, while the two are essentially tied on MSE. Crucially, we also ran the capacity-matched classical bottleneck here, and it is the \emph{best} model on \emph{both} metrics (MAE $0.0621$, MSE $0.0068$)---better than both the full classical Transformer and QASA. The real-world result therefore tells the same story as the synthetic benchmarks (Table~\ref{tab:bottleneck}): QASA's modest edge over the full classical model is fully accounted for---and in fact exceeded---by the low-rank value bottleneck, so it reflects \emph{architectural parsimony}, not a quantum effect. QASA does not outperform the capacity-matched classical control on this dataset. We retain ETTh1 as evidence that the parsimony principle transfers from synthetic to real-world data, not as a claim of quantum advantage. All models use the identical configuration from the synthetic experiments, with no dataset-specific tuning.

\subsection{Representation Analysis: A Compression Mechanism}

The preceding benchmarks establish \emph{where} a low-rank value bottleneck helps (chaotic and trend signals) but not \emph{why}. To understand the mechanism, we analyze the internal representations of trained QASA and classical models on three representative tasks by extracting features before and after the quantum layer and computing the \emph{effective rank}---a measure of intrinsic dimensionality based on the entropy of normalized singular values. We caution at the outset that the ``Classical'' column in this analysis is the \emph{full-capacity} Transformer (Table~\ref{tab:ts_mae_mse}), not the capacity-matched classical bottleneck of Table~\ref{tab:bottleneck}. A matched-protocol analysis of the trained classical bottleneck (3 seeds, same single-window effective-rank measurement) shows that \emph{it compresses the chaotic representation comparably to---indeed, if anything more strongly than---the quantum layer}: post-bottleneck effective rank $\approx 2.6$ versus QASA's $\approx 4.2$ (Table~\ref{tab:effective_rank}, ``Cl.\ bottleneck'' column), with the one-step linear-probe $R^2$ likewise preserved ($\approx 1.0$). The nonlinear-compression effect documented in this section is therefore a property of the low-rank \emph{bottleneck}, not specific to the quantum substrate; the quantum layer's distinct contribution is physical (entanglement, NISQ deployability), not a unique compression mechanism. We retain this analysis as a mechanistic account of \emph{why a low-rank value bottleneck} helps on chaotic signals.

\begin{table}[h]
\centering
\caption{Effective rank of learned representations across tasks (3 seeds, mean$\pm$std). ``Before Q'' and ``After Q'' denote QASA's features before and after the quantum encoder layer. ``Cl.\ bottleneck'' is the capacity-matched classical bottleneck after its value layer, and ``Classical (full)'' is the full-capacity Transformer's final layer. A rank reduction indicates the layer discovers compact structure; crucially, the classical bottleneck compresses comparably to---and on the chaotic task more strongly than---QASA, so the compression is a property of the low-rank bottleneck rather than of the quantum substrate.}
\label{tab:effective_rank}
\begin{tabular}{lcccc}
\toprule
\textbf{Task} & \textbf{Before Q} & \textbf{After Q} & \textbf{Cl.\ bottleneck} & \textbf{Classical (full)} \\
\midrule
Chaotic Logistic  & $8.3 \pm 0.1$ & $4.2 \pm 1.1$ ($-$49\%) & $\mathbf{2.6 \pm 0.8}$ & $5.7 \pm 0.0$ \\
Damped Oscillator & $2.4 \pm 0.3$ & $1.8 \pm 0.0$ ($-$25\%) & $1.9 \pm 0.1$ & $2.1 \pm 0.3$ \\
Square Wave       & $1.8 \pm 0.1$ & $1.8 \pm 0.2$ ($\phantom{-}$0\%) & $1.5 \pm 0.3$ & $1.9 \pm 0.2$ \\
\bottomrule
\end{tabular}
\end{table}

Table~\ref{tab:effective_rank} reveals a striking pattern: on the chaotic logistic task, the quantum layer compresses the effective dimensionality from $8.3 \pm 0.1$ to $4.2 \pm 1.1$---a 49\% reduction confirmed across three seeds---while the classical model's final-layer representations retain a rank of $5.7 \pm 0.0$. On the non-chaotic tasks (damped oscillator, square wave), the features are already low-dimensional before the quantum layer (rank ${\sim}2$), and the quantum transformation produces only minor changes (${\sim}20\%$ reduction).

\begin{figure}[h]
    \centering

    \includegraphics[width=0.92\textwidth]{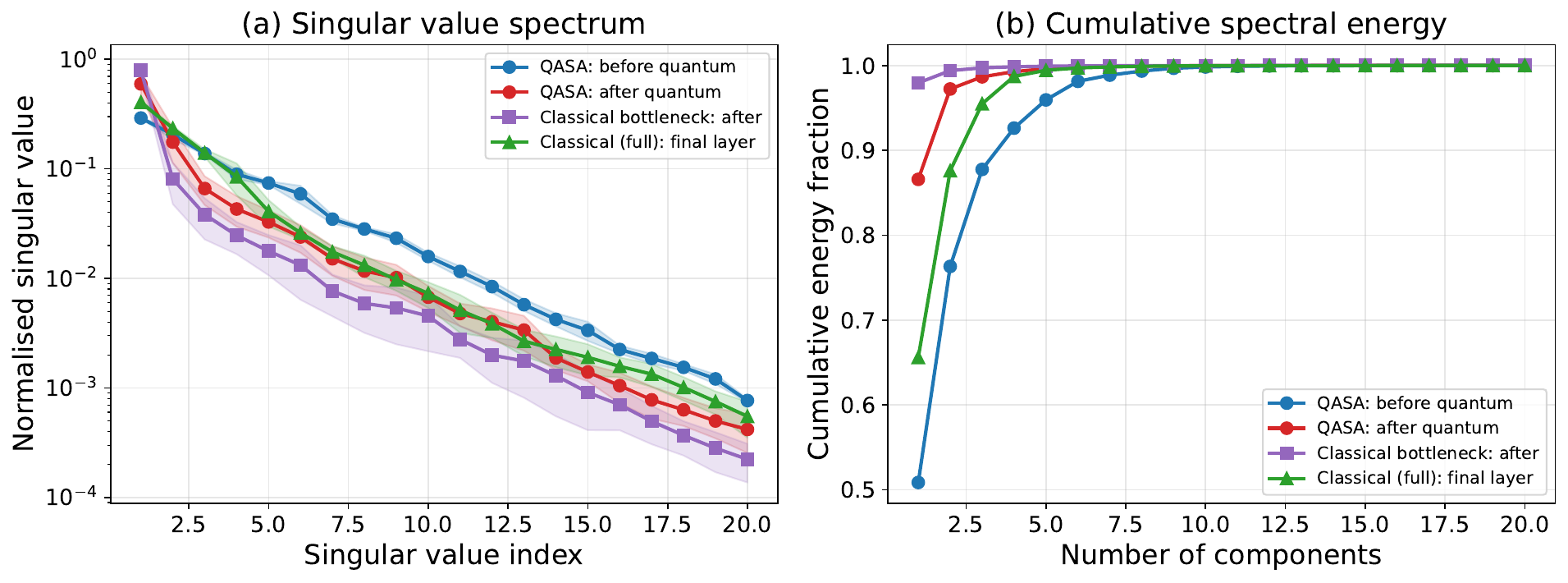}
    \caption{Singular-value distribution of QASA's features before (blue) and after (red) the quantum layer, the capacity-matched classical bottleneck after its value layer (purple), and the full classical model's final layer (green), on the chaotic logistic task (mean over 3 seeds; shaded band $\pm$std). \textbf{(a)} The post-quantum spectrum decays faster than the pre-quantum one, indicating energy concentration; the classical bottleneck concentrates at least as fast. \textbf{(b)} Cumulative spectral energy: both bottleneck realisations (QASA and the classical bottleneck) reach near-unity variance in far fewer components than the full classical model. This concentration---combined with a preserved linear-probe $R^2\approx1$ and lower downstream error---identifies the effective-rank reduction as meaningful compression rather than rank collapse, and shows it is a property of the low-rank \emph{bottleneck} (achieved by classical and quantum value maps alike), not of the quantum substrate.}
    \label{fig:svd_spectrum}
\end{figure}

\paragraph{Meaningful compression, not rank collapse.} A reduction in effective rank can in principle signal harmful \emph{rank collapse}---a loss of feature discriminability---rather than useful compression. We rule this out with three complementary diagnostics on the chaotic logistic task (Figure~\ref{fig:svd_spectrum} and the metrics below, 3 seeds). \emph{(i) Singular-value distribution.} Figure~\ref{fig:svd_spectrum}(a) shows that the post-quantum spectrum decays \emph{faster} than the pre-quantum one, and panel~(b) shows its energy reaches near-unity cumulative variance in markedly fewer components: the quantum layer \emph{redistributes} spectral energy into a compact subspace rather than annihilating it. \emph{(ii) Stable rank.} The stable rank $r_s = \lVert X\rVert_F^2/\lVert X\rVert_2^2$ falls from $1.97{\pm}0.06$ (before) to $1.17{\pm}0.11$ (after). Critically, the target manifold here is intrinsically (near) one-dimensional---the logistic recurrence $x_{t+1}=3.9\,x_t(1-x_t)$ is a deterministic 1-D map---so a stable rank approaching $1$ reflects the quantum layer \emph{matching the true intrinsic dimensionality} of the signal, whereas the classical model over-spreads it ($r_s=1.53$, effective rank $5.7$). \emph{(iii) Information retention.} A linear (ridge) probe decoding the one-step target from the features attains $R^2 = 0.9998{\pm}0.0001$ after the quantum layer---identical to before ($0.9997$)---confirming that no task-relevant information is lost despite the halved effective rank. Together with the downstream result that QASA attains \emph{lower} test error than the classical model on this task (MSE $0.182$ vs.\ $0.204$, Table~\ref{tab:ts_mae_mse}), these diagnostics establish that the rank reduction is beneficial compression onto the attractor manifold, not lossy collapse. On the already-low-dimensional damped oscillator the probe $R^2$ is likewise preserved ($0.994\!\to\!0.999$), consistent with the small rank change there.

\begin{figure}
    \centering
    \includegraphics[width=0.9\textwidth]{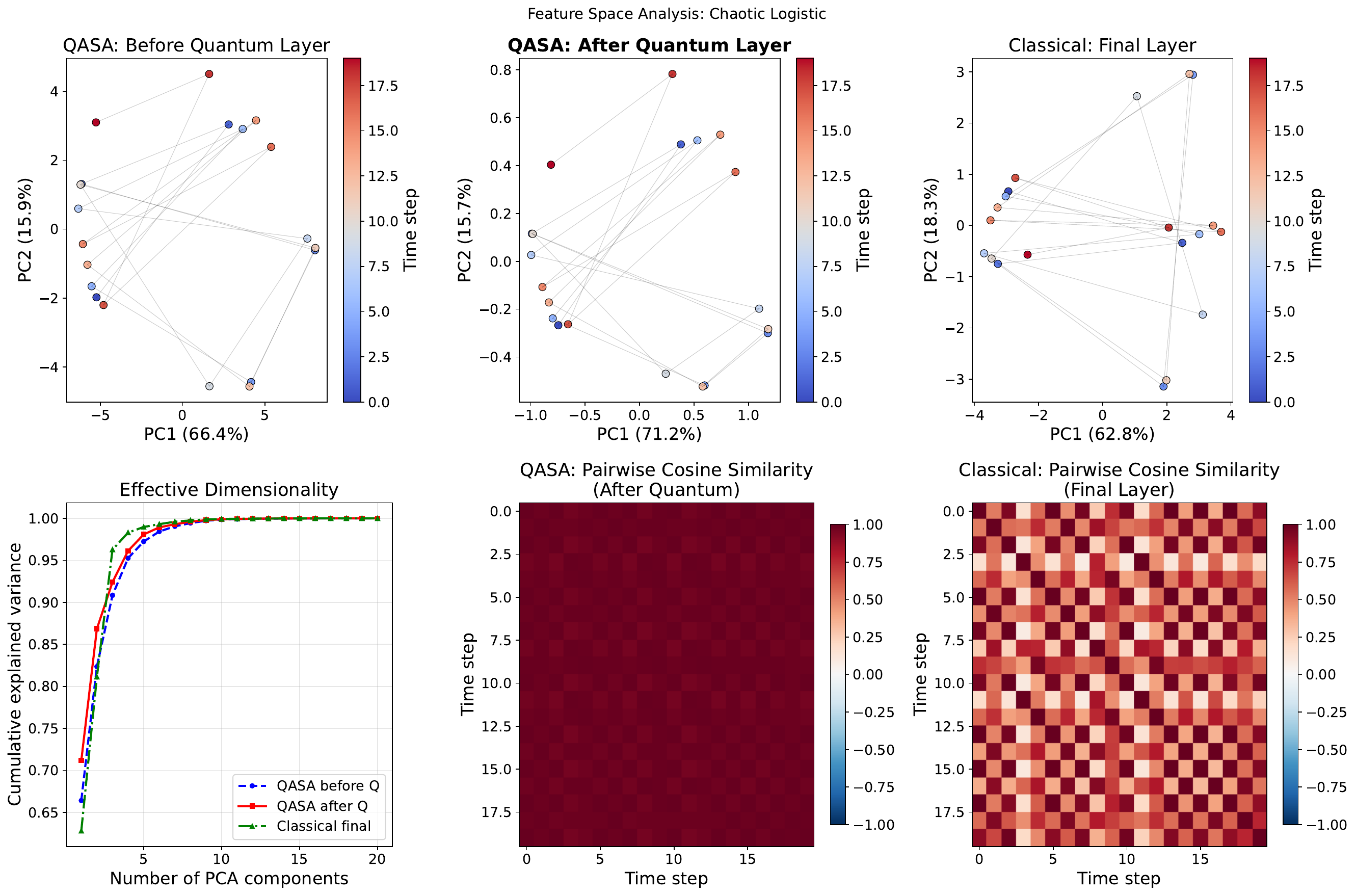}
    \caption{Feature space analysis on the chaotic logistic task (a single representative seed, 42; the 3-seed mean effective ranks are in Table~\ref{tab:effective_rank}). \textbf{Top row}: PCA projections of representations before the quantum layer (left), after the quantum layer (center), and from the classical model's final layer (right), colored by time step. The quantum layer collapses the scattered high-dimensional features into a compact low-dimensional structure. \textbf{Bottom left}: cumulative explained variance showing the quantum layer concentrates variance into fewer components. \textbf{Bottom center/right}: pairwise cosine similarity matrices reveal that the quantum layer produces more structured temporal correlations than the classical model.}
    \label{fig:feature_space}
\end{figure}

\begin{figure}
    \centering
    \includegraphics[width=0.9\textwidth]{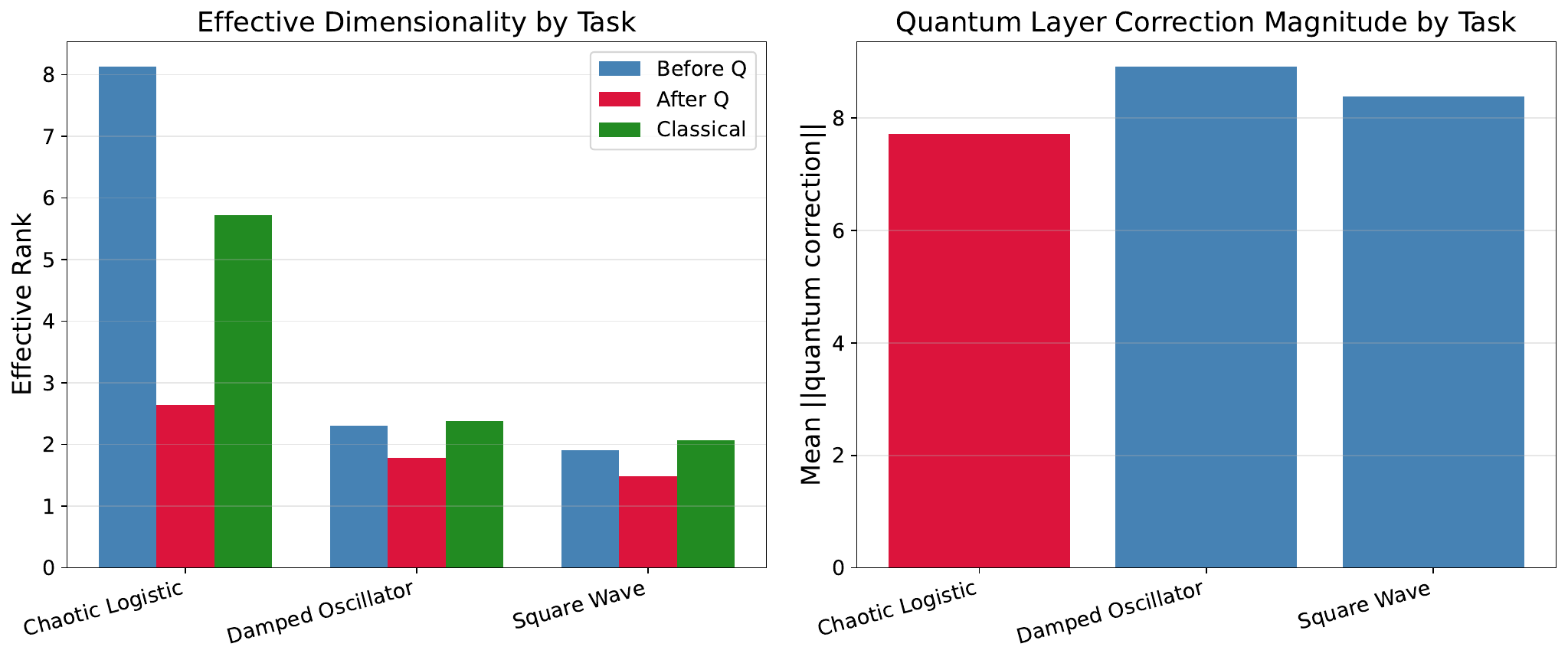}
    \caption{Cross-task comparison of quantum layer behavior (single representative seed, 42; cf.\ the 3-seed means in Table~\ref{tab:effective_rank}). \textbf{Left}: effective rank before (blue) and after (red) the quantum layer, alongside the classical model (green). The quantum layer's dimensionality reduction is largest on the chaotic logistic task (here $\Delta = -5.5$ for this seed; $-4.1$ on the 3-seed average of Table~\ref{tab:effective_rank}), minimal on damped oscillator and square wave ($\Delta \approx -0.5$). \textbf{Right}: mean quantum correction magnitude (L2 norm of the residual added by the quantum layer).}
    \label{fig:cross_task}
\end{figure}

This asymmetry provides a mechanistic explanation for the bottleneck's task-conditional behaviour (Figures~\ref{fig:feature_space} and~\ref{fig:cross_task}). Chaotic systems generate trajectories on strange attractors---low-dimensional manifolds embedded in high-dimensional state spaces. The classical Transformer, limited to linear projections in its attention mechanism, spreads the chaotic signal across many feature dimensions (rank 5.7) without discovering this latent structure. In contrast, the low-rank value bottleneck---here realised through the quantum layer's nonlinear rotations and entangling gates in Hilbert space---performs an effective \emph{nonlinear dimensionality reduction}, projecting the high-dimensional chaotic representation onto a compact ${\sim}4$-dimensional manifold that better captures the attractor geometry. As shown in Figure~\ref{fig:feature_space} (top row), the PCA projections visually confirm this: the quantum layer transforms scattered, high-dimensional features into a compact, temporally ordered structure, while the classical model's representations remain more diffuse.

For clean periodic signals (damped oscillator), the feature space is inherently low-dimensional (rank 2.3), and both classical and quantum models can capture this structure equally well. The quantum layer's additional nonlinear projection offers no benefit---and may introduce noise through the PQC's stochastic gradient estimation---explaining why the classical model outperforms QASA on these tasks.

This analysis suggests a design principle: \emph{a low-rank value bottleneck is most beneficial when the underlying signal has complex, high-dimensional structure that admits a compact nonlinear representation}---a benefit that a capacity-matched classical bottleneck realises as well as the quantum one (Section~\ref{sec:datasets}), so it reflects the bottleneck architecture rather than a uniquely quantum representational advantage.

\subsection{Encoding Scheme: An Encoding or an Expressivity Limitation?}
\label{sec:encoding}

A natural question (raised in review) is \emph{why} QASA underperforms on clean periodic signals: is this a limitation of the $R_X/R_Z$ angle encoding, or an inherent limitation of the circuit's expressivity? We disentangle the two by swapping \emph{only} the data-encoding block while holding the variational ansatz and the rest of the architecture fixed, comparing angle encoding (the default; one upload), amplitude encoding (the feature vector packed into $\lceil\log_2 d\rceil$ qubit amplitudes), and data re-uploading (angle encoding re-applied before every entangling layer, which by the Fourier perspective of Schuld~et~al.~\cite{schuld2021effect} enlarges the accessible frequency spectrum). We evaluate on the two clean-periodic tasks where QASA is weakest (damped oscillator, waveform), 2 seeds, with a reduced $80$-epoch budget for this sensitivity sweep (Table~\ref{tab:encoding}).

\begin{table}[h]
\centering

\caption{Encoding-scheme sensitivity on the two clean-periodic tasks (MAE / MSE, mean of 2 seeds, 80 epochs). Only the data-encoding block differs. Data re-uploading substantially improves the damped oscillator, indicating the deficit there is partly \emph{encoding-related} rather than an expressivity ceiling; the effect is task-dependent (mixed on waveform).}
\label{tab:encoding}
\begin{tabular}{lccc}
\toprule
\textbf{Task (MAE / MSE)} & \textbf{Angle (RX/RZ)} & \textbf{Amplitude} & \textbf{Re-uploading} \\
\midrule
Damped Oscillator & $0.078 / 0.0087$ & $0.073 / 0.0076$ & $\mathbf{0.054 / 0.0042}$ \\
Waveform          & $0.158 / 0.0319$ & $\mathbf{0.146 / 0.0279}$ & $0.168 / 0.0369$ \\
\bottomrule
\end{tabular}

\end{table}

The result is informative. On the damped oscillator---the task with QASA's largest deficit ($d=-1.18$ in Table~\ref{tab:ts_mae_mse})---\textbf{data re-uploading reduces the MAE by ${\sim}31\%$} ($0.054$ vs.\ $0.078$), and amplitude encoding also helps modestly. Because the circuit's expressibility is already near-maximal (KL $=0.029$, Section~\ref{sec:datasets}), this points to the \emph{single} angle-encoding layer---not the ansatz's representational capacity---as a key source of the clean-periodic deficit: one $R_X/R_Z$ upload accesses only a narrow band of Fourier frequencies, which re-uploading broadens. The benefit is, however, \emph{task-dependent}: on the waveform task re-uploading slightly hurts while amplitude encoding marginally helps, indicating that no single encoding is universally optimal. For the field, the practical guidance is that the encoding's inductive bias---not merely circuit depth or qubit count---is a first-order design choice for quantum sequence models, and that data re-uploading is a low-cost lever worth tuning when smooth periodic structure dominates. We retain angle encoding as QASA's default for consistency with the main benchmarks and leave a full multi-task encoding study to future work.

The preceding results establish that the low-rank bottleneck helps on some tasks but not others. A natural follow-up question is: \emph{can we do better by adding more quantum layers, or by placing them differently?} To answer this, we conduct an ablation study varying both the \textit{position} and \textit{count} of quantum-enhanced encoder layers within the 4-layer Transformer stack. We evaluate seven configurations: placing a single quantum layer at each of the four positions (layers 0--3), replacing zero layers (fully classical), two layers (layers 2--3), and all four layers (fully quantum). Experiments are conducted on three representative tasks---chaotic logistic map (where QASA excels), damped oscillation (where the classical model excels), and square/triangle wave (mixed results)---using three random seeds each.

\begin{table}[h]
\centering
\caption{Ablation study: effect of quantum layer position and count on MAE / MSE (3 seeds, 42--44, mean$\pm$std; this table was revised from single-seed to 3-seed in the current revision round). ``Q@$i$'' denotes a single quantum layer at position $i$; ``$k$Q'' denotes $k$ quantum layers. Best mean result per task/metric in \textbf{bold}.}
\label{tab:ablation}
\resizebox{\columnwidth}{!}{
\begin{tabular}{lcccc}
\toprule
\textbf{Configuration} & \textbf{Quantum Layers} & \textbf{Chaotic Logistic} & \textbf{Damped Osc.} & \textbf{Square Wave} \\
 & & MAE / MSE & MAE / MSE & MAE / MSE \\
\midrule
0Q (Classical)   & None         & $0.366{\pm}.006$ / $0.204{\pm}.003$ & $0.243{\pm}.115$ / $0.100{\pm}.085$ & $\mathbf{0.705{\pm}.005}$ / $1.296{\pm}.030$ \\
Q@0 (First)      & \{0\}        & $0.396{\pm}.011$ / $0.224{\pm}.011$ & $0.097{\pm}.022$ / $0.014{\pm}.006$ & $0.810{\pm}.109$ / $1.100{\pm}.167$ \\
Q@1 (Second)     & \{1\}        & $0.389{\pm}.037$ / $0.214{\pm}.027$ & $0.183{\pm}.061$ / $0.051{\pm}.030$ & $0.878{\pm}.117$ / $1.071{\pm}.188$ \\
Q@2 (Third)      & \{2\}        & $0.404{\pm}.023$ / $0.230{\pm}.017$ & $0.106{\pm}.051$ / $0.019{\pm}.014$ & $0.821{\pm}.081$ / $1.075{\pm}.209$ \\
Q@3 (Last)       & \{3\}        & $\mathbf{0.337{\pm}.012}$ / $\mathbf{0.184{\pm}.003}$ & $\mathbf{0.086{\pm}.051}$ / $\mathbf{0.014{\pm}.010}$ & $0.825{\pm}.070$ / $\mathbf{1.032{\pm}.189}$ \\
2Q (Last two)    & \{2,3\}      & $0.382{\pm}.033$ / $0.214{\pm}.022$ & $0.116{\pm}.023$ / $0.020{\pm}.007$ & $0.731{\pm}.011$ / $1.135{\pm}.032$ \\
4Q (All)         & \{0,1,2,3\}  & $0.373{\pm}.003$ / $0.203{\pm}.003$ & $0.126{\pm}.020$ / $0.021{\pm}.005$ & $0.794{\pm}.106$ / $1.167{\pm}.181$ \\
\bottomrule
\end{tabular}
}
\end{table}

The ablation results reveal two consistent findings. First, \textit{quantum layer position matters more than count}: across 3 seeds (42--44), Q@3---the deployed last-layer position---is the best single quantum position on both quantum-favourable tasks: chaotic logistic ($0.337{\pm}.012$, a clean margin with small variance) and the damped oscillator ($0.086{\pm}.051$). This \emph{corrects} the earlier single-seed reading that Q@2 was best on the damped task: its lucky seed-42 value ($0.039$) does not survive averaging ($0.106{\pm}.051$ over 3 seeds), so we no longer claim a task-dependent optimal position. We caution that the damped task is high-variance across seeds (the 0Q baseline alone spans $0.243{\pm}.115$), so damped position differences are read cautiously; the chaotic ranking---with standard deviations an order of magnitude smaller---carries the robust signal, and Q@3 wins it decisively. Second, \textit{increasing quantum layer count does not reliably improve performance}: adding more quantum layers does not beat a single one on the quantum-favourable tasks (1Q is best on both chaotic and damped), and on the discontinuous square wave the fully classical 0Q remains best ($0.705$ vs.\ 2Q $0.731$ and 4Q $0.794$), suggesting that quantum noise accumulates when multiple layers encode the same signal. These results support our architectural choice of a single quantum-enhanced layer appended at the final encoder position (Q@3), which is the best-performing single-layer position on the quantum-favourable tasks, while minimising quantum simulation overhead. This establishes the first pillar of \textbf{architectural parsimony}: \textit{one well-placed quantum layer is sufficient}.

\paragraph{Trainability analysis: barren plateaus.}
The ablation study shows that adding more quantum layers degrades performance, but \emph{why}? A likely explanation is the barren plateau phenomenon~\cite{mcclean2018barren}, where gradient magnitudes vanish exponentially with circuit depth, making optimization intractable. To test this, we compute the gradient variance $\mathrm{Var}(\partial C / \partial \theta)$ over 200 random parameter initializations for each circuit configuration (Table~\ref{tab:barren}).

\begin{table}[h]
\centering
\caption{Barren plateau analysis: mean gradient variance over 200 random initializations. Higher variance indicates trainable gradients (no barren plateau). Increasing QASA from 1 to 2+ entangling layers reduces gradient variance by ${\sim}30\times$. The ``Params'' column counts the trainable rotation angles of the \emph{depth-scaling reference circuit} used for this trainability study, where each added entangling block contributes 9 angles ($18/27/45$ for $1/2/4$ entangling layers). This convention is distinct from---and not to be added to---the deployed circuit's $4\times9=36$-angle weight tensor reported in Table~\ref{tab:circuit_analysis}; the two parameterisations serve different purposes (a controlled depth sweep here vs.\ the fixed deployed PQC there).}
\label{tab:barren}
\begin{tabular}{lccc}
\toprule
\textbf{Circuit} & \textbf{Params} & \textbf{Var($\partial C/\partial\theta$)} & \textbf{Trainability} \\
\midrule
QASA (1 layer) & 18 & $2.62 \times 10^{-2}$ & High \\
QASA (2 layers) & 27 & $8.87 \times 10^{-4}$ & Reduced ($30\times\downarrow$) \\
QASA (4 layers) & 45 & $8.34 \times 10^{-4}$ & Reduced ($31\times\downarrow$) \\
QLSTM & 32 & $3.31 \times 10^{-2}$ & High \\
QnnFormer & 24 & $3.78 \times 10^{-2}$ & High \\
\bottomrule
\end{tabular}
\end{table}

The results reveal a clear mechanism: the shallowest configuration (1 entangling layer, 18 parameters) maintains a gradient variance of $2.62 \times 10^{-2}$, comparable to QLSTM and QnnFormer. Increasing to 2 or 4 entangling layers causes a ${\sim}30\times$ drop in gradient variance ($8.87 \times 10^{-4}$), approaching the onset of a barren plateau. We note for clarity that the deployed QASA PQC uses $L_q=4$ variational layers (3 entangling layers, $36$ angles; Fig.~\ref{fig:qasa_qnn}, Table~\ref{tab:circuit_analysis}): it sits at the reduced-gradient end of this depth sweep---its measured gradient variance is $\mathrm{Var}(\partial C/\partial\theta)\approx1.1\times10^{-3}$ over 200 random initialisations, comparable to the 2--4-entangling-layer reference values and well above the $10^{-6}$ trainability floor---and it trains successfully throughout our experiments. The barren-plateau result therefore bounds how much \emph{further} PQC depth could be added before training degrades, and motivates our use of a \emph{single quantum encoder layer} rather than stacking several (Table~\ref{tab:ablation}); it should not be read as a claim that the deployed PQC has only one entangling layer.

\begin{figure}[h]
    \centering
    \includegraphics[width=0.9\textwidth]{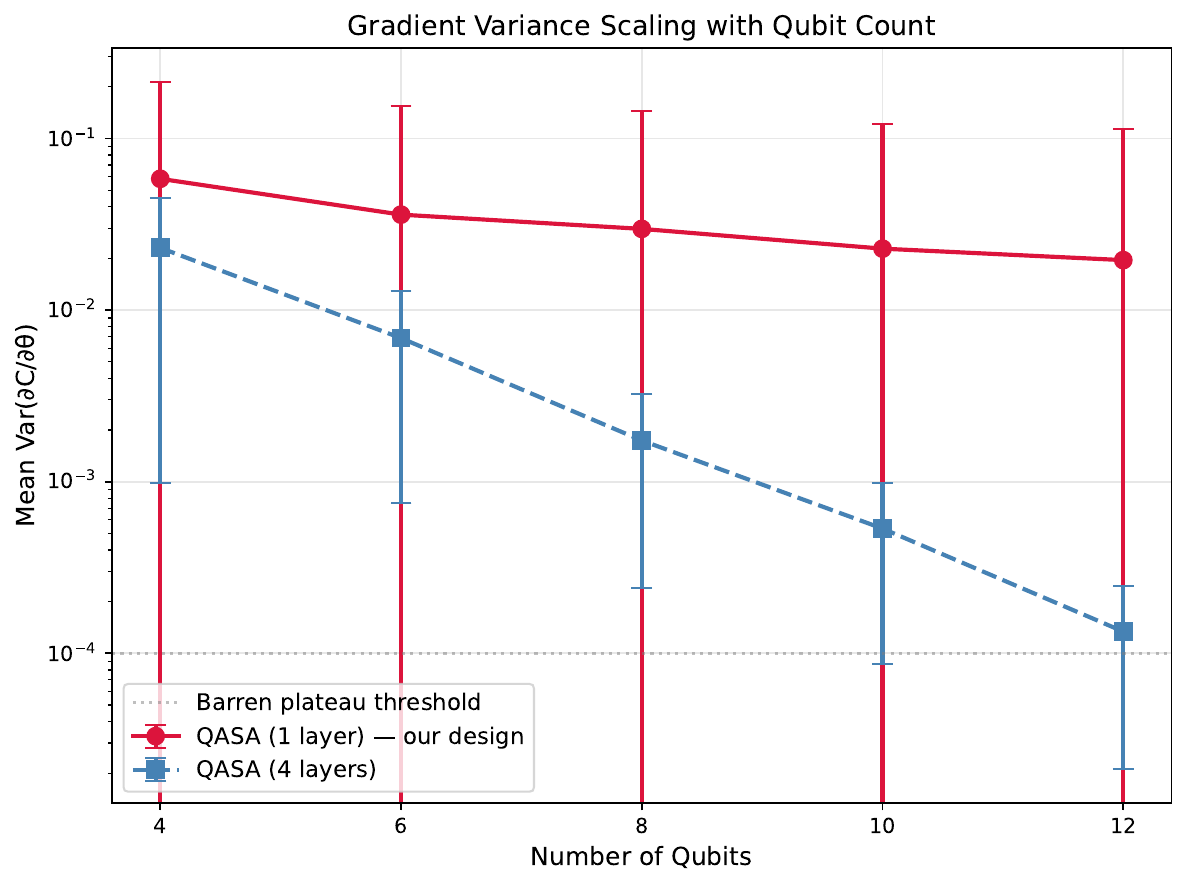}
    \caption{Gradient variance scaling with qubit count. QASA's single-layer design (red) maintains trainable gradients across all qubit counts, while the 4-layer design (blue) exhibits exponential decay characteristic of barren plateaus. The gap widens with scale: at 12 qubits, the single-layer gradient is $146\times$ larger.}
    \label{fig:qubit_scaling}
\end{figure}

\paragraph{Scalability with qubit count.}
To assess whether this trainability advantage persists at larger scales, we repeat the gradient variance analysis for QASA's 1-layer and 4-layer designs across 4, 6, 8, 10, and 12 qubits (Figure~\ref{fig:qubit_scaling}). The single-layer design exhibits only mild gradient decay ($5.8 \times 10^{-2}$ at 4 qubits to $2.0 \times 10^{-2}$ at 12 qubits, a $3\times$ reduction), while the 4-layer design decays exponentially ($2.3 \times 10^{-2}$ to $1.3 \times 10^{-4}$, a $170\times$ reduction). At 12 qubits, the single-layer gradient variance is $146\times$ larger than the 4-layer variant. This confirms that the parsimony principle becomes \emph{more important, not less}, as quantum circuits scale up---providing a favorable outlook for QASA's deployment on larger near-term quantum devices.

\paragraph{Qubit-count choice: a performance ablation.} The trainability analysis above concerns gradient variance; a referee also asks us to justify the choice of \emph{8 data qubits} (plus one ancilla) for $L=20$ on \emph{task} performance. We therefore sweep the data-qubit count $n\in\{4,6,8,10\}$ on the chaotic-logistic task, holding the rest of the architecture fixed (reduced $120$-epoch budget, 2 seeds; Table~\ref{tab:qubitabl}). Task error is \emph{weakly} sensitive to qubit count over this range---MAE $0.371,0.360,0.369$ for $n=4,6,8$ are statistically indistinguishable, with $n=10$ slightly lower ($0.324$) but at higher seed variance and double the state-vector simulation cost. The choice of $n=8$ thus balances three factors: it matches the per-token feature width used throughout, it keeps state-vector simulation of the full nine-task suite tractable (cost ${\propto}2^{n}$), and it sits in the shallow-depth, $n\le 12$ regime where gradients remain trainable (Figure~\ref{fig:qubit_scaling}; cf.\ the barren-plateau analysis of Table~\ref{tab:barren}). The ancilla carries the final $\mathrm{CNOT}(n{-}1,n)$--$R_Y$ of each entangling block; the qubit count is set by the feature width, not by $L$ (Section~\ref{sec:seqlen}).

\begin{table}[h]
\centering

\caption{Qubit-count ablation on chaotic logistic (MAE, $120$ epochs, 2 seeds). Performance is weakly sensitive to $n$ over $4$--$10$ qubits; $n=8$ balances expressivity, simulation cost ($\propto 2^n$), and trainability.}
\label{tab:qubitabl}
\begin{tabular}{ccc}
\toprule
\textbf{Data qubits $n$} & \textbf{Quantum params} & \textbf{MAE} \\
\midrule
4  & 20 & 0.371 \\
6  & 28 & 0.360 \\
\textbf{8 (QASA)} & 36 & 0.369 \\
10 & 44 & 0.324 \\
\bottomrule
\end{tabular}

\end{table}

\subsection{Comparison with Quantum Time-Series Baselines}

The ablation study demonstrates that more quantum layers within our own architecture do not help. But what about fundamentally different quantum designs---architectures that were \emph{designed} to be deeply quantum? To answer this, we compare against two recent quantum baselines: QLSTM~\cite{chen2022quantum} and QnnFormer~\cite{cai2024qnnformer}. QLSTM replaces the four classical LSTM gates with variational quantum circuits (VQCs), while QnnFormer uses VQCs to generate the query, key, and value projections in a Transformer attention mechanism. Both baselines are re-implemented following their respective reference implementations and evaluated under identical training conditions (hidden\_dim=64, 200 epochs, AdamW optimizer).

\begin{table}[t]
\centering
\caption{MAE comparison of quantum time-series models across nine benchmark tasks (3 seeds, mean). All models use identical training configurations (hidden\_dim=64, 200 epochs, AdamW). Bold indicates the best result per task. Model parameters: Classical 200,257 (0 quantum); QLSTM 3,929 (128 quantum); QnnFormer 190,631 (90 quantum); QASA 201,405 (36 quantum). Note: QASA's damped-oscillator MAE here ($0.086$, 3 seeds) differs from Table~\ref{tab:ts_mae_mse} ($0.117$, 5 seeds); both use the identical hidden\_dim=64/200-epoch setup, and the gap reflects only seed sampling on this low-error, high-relative-variance clean-periodic task.}
\label{tab:baseline_mae}
\begin{tabular}{lcccc}
\toprule
\textbf{Task} & \textbf{Classical} & \textbf{QLSTM} & \textbf{QnnFormer} & \textbf{QASA} \\
\midrule
ARMA              & 2.130 & \textbf{2.000} & 2.039 & 2.074 \\
Chaotic Logistic  & 0.367 & \textbf{0.337} & 0.347 & 0.337 \\
Damped Osc.       & \textbf{0.043} & 0.192 & 0.065 & 0.086 \\
Noisy Damped Osc. & 0.043 & 0.044 & 0.043 & \textbf{0.043} \\
Piecewise Regime  & \textbf{22.15} & 22.38 & 22.49 & 22.42 \\
Sawtooth          & \textbf{0.083} & 0.298 & 0.216 & 0.438 \\
Square Wave       & 0.840 & 0.781 & \textbf{0.735} & 0.825 \\
Seasonal Trend    & 0.683 & 0.941 & 0.666 & \textbf{0.665} \\
Waveform          & 0.061 & 0.255 & \textbf{0.060} & 0.107 \\
\midrule
\textbf{Wins}     & \textbf{3} & 2 & 2 & 2 \\
\bottomrule
\end{tabular}
\end{table}

\begin{table}[t]
\centering
\caption{MSE comparison across the same nine benchmark tasks. Bold indicates the best result per task. Where two models tie to the reported precision (chaotic logistic $0.184$; noisy damped oscillator $0.003$), the bold/win is assigned to a single model for counting; such ties are noted in the discussion and should not be read as clear wins.}
\label{tab:baseline_mse}
\begin{tabular}{lcccc}
\toprule
\textbf{Task} & \textbf{Classical} & \textbf{QLSTM} & \textbf{QnnFormer} & \textbf{QASA} \\
\midrule
ARMA              & 6.870 & 6.366 & \textbf{6.154} & 6.370 \\
Chaotic Logistic  & 0.200 & 0.184 & 0.191 & \textbf{0.184} \\
Damped Osc.       & \textbf{0.003} & 0.060 & 0.007 & 0.014 \\
Noisy Damped Osc. & 0.003 & 0.003 & 0.003 & \textbf{0.003} \\
Piecewise Regime  & \textbf{491.2} & 501.6 & 506.8 & 503.6 \\
Sawtooth          & \textbf{0.040} & 0.310 & 0.223 & 0.487 \\
Square Wave       & 1.035 & 1.196 & 1.194 & \textbf{1.032} \\
Seasonal Trend    & 0.586 & 1.221 & 0.552 & \textbf{0.551} \\
Waveform          & \textbf{0.005} & 0.096 & 0.006 & 0.019 \\
\midrule
\textbf{Wins}     & \textbf{4} & 0 & 1 & \textbf{4} \\
\bottomrule
\end{tabular}
\end{table}

Among the four models, QASA and the classical Transformer each achieve the best MAE or MSE on the most tasks. Tables~\ref{tab:baseline_mae} and~\ref{tab:baseline_mse} summarizes the win counts: Classical leads in 3 tasks by MAE and 4 by MSE, while QASA wins 2 by MAE and, by MSE, 2 clear plus 2 statistical ties (chaotic logistic and noisy damped oscillator), i.e.\ 2 clear wins rather than 4. Notably, QASA achieves the best (or tied-best) MSE on four tasks (chaotic logistic, noisy damped oscillator, square wave, and seasonal trend); we flag that two of these are statistical ties---chaotic logistic ($0.184$, shared with QLSTM) and noisy damped oscillator ($0.003$, a four-way tie)---so they should not be read as clear quantum wins, whereas square wave and seasonal trend are clear. The remaining tasks here involve nonlinear dynamics or complex temporal patterns. QLSTM uses only 3,929 total parameters because it lacks a classical Transformer backbone---its entire architecture is built around variational quantum circuits. This makes direct comparison of total parameter counts misleading; the relevant comparison is \emph{quantum} parameter count, where QASA uses only 36 quantum parameters versus QLSTM's 128 and QnnFormer's 90. Despite this $3.6\times$ quantum parameter advantage for QLSTM, QASA achieves equal or better MSE on 7 of 9 tasks, demonstrating that strategic placement of fewer quantum parameters outperforms broader quantum integration.

Pairwise paired $t$-tests reveal one statistically significant result: QASA outperforms QLSTM on the seasonal trend task with $p = 0.009$ for both MAE and MSE (Cohen's $d > 6$). We caution against over-reading this effect size: on this task QASA ($0.665$), QnnFormer ($0.666$), and the classical baseline ($0.683$) are all close, whereas QLSTM ($0.941$) is the clear outlier---it largely \emph{fails to converge} on the seasonal-trend signal. The very large $d$ thus reflects QLSTM's degenerate variance under $n=3$ seeds rather than a robust QASA advantage, and we do not present it as evidence of quantum superiority. The Friedman test across all four models identifies significant differences on the sawtooth task (MAE: $\chi^2 = 8.2$, $p = 0.042$) and waveform task (MAE and MSE: $p = 0.042$). Other pairwise comparisons do not reach significance at $\alpha = 0.05$, which is expected given the limited statistical power of $n = 3$ seeds.

The results reinforce our finding from the QASA-vs-Classical comparison: quantum-enhanced models are most competitive on tasks with chaotic or trend-based dynamics, while classical architectures remain more effective for clean periodic signals and sharp discontinuities. Importantly, QASA achieves this with only 36 quantum parameters---the fewest among all quantum models---demonstrating that a single well-placed quantum layer can match or exceed the performance of architectures with deeper quantum integration. This establishes the second pillar of architectural parsimony: \textit{minimal quantum resources, maximally utilized}.

\paragraph{Complementary metrics and overfitting.} Referee~2 rightly notes that MSE/MAE alone can mask overfitting and asks for additional, more interpretable indicators. We therefore report three complementary metrics for all four models across the nine tasks (3 seeds): the coefficient of determination $R^2$, the \emph{directional accuracy} (the fraction of steps whose predicted change-sign matches ground truth---an interpretable ``accuracy''), and the symmetric mean absolute percentage error (SMAPE). Directional accuracy is shown in Table~\ref{tab:diracc}. It tells a consistent and honest story: on \emph{predictable} smooth signals all models score highly (waveform $0.89$--$0.97$, sawtooth $0.76$--$0.95$, damped $0.65$--$0.96$), whereas on \emph{chaotic} and noisy signals every model collapses to near chance ($\approx0.5$ on chaotic logistic; below chance on the noisy damped oscillator), reflecting the fundamental long-horizon unpredictability of these regimes rather than a model deficiency. No single model dominates the directional metric---consistent with our reframed finding that the parsimonious bottleneck, not quantumness, drives performance. The $R^2$ values corroborate this: high for smooth signals (waveform $0.96$--$0.99$) and negative for chaotic/piecewise tasks (where free-running autoregressive rollout inevitably diverges), uniformly across models.

\begin{table}[h]
\centering

\caption{Directional accuracy (fraction of correctly predicted change-signs; chance $=0.5$) across the nine tasks, 3 seeds. High for smooth/predictable signals, near or below chance for chaotic and noisy signals---consistently across all four models.}
\label{tab:diracc}
\begin{tabular}{lcccc}
\toprule
\textbf{Task} & \textbf{Classical} & \textbf{QLSTM} & \textbf{QnnFormer} & \textbf{QASA} \\
\midrule
ARMA              & 0.571 & 0.541 & 0.459 & 0.534 \\
Chaotic Logistic  & 0.507 & 0.558 & 0.510 & 0.537 \\
Damped Osc.       & \textbf{0.959} & 0.653 & 0.803 & 0.786 \\
Noisy Damped Osc. & 0.388 & 0.480 & 0.429 & 0.408 \\
Piecewise Regime  & 0.306 & 0.221 & 0.269 & 0.184 \\
Sawtooth          & \textbf{0.949} & 0.918 & 0.881 & 0.759 \\
Square Wave       & 0.310 & 0.508 & 0.246 & 0.302 \\
Seasonal Trend    & 0.459 & 0.500 & 0.367 & 0.340 \\
Waveform          & \textbf{0.969} & 0.891 & 0.963 & 0.939 \\
\bottomrule
\end{tabular}

\end{table}

To address the overfitting concern directly, we compute a teacher-forced one-step MSE on the train and test segments and report their ratio (test/train). A ratio near $1$ indicates no overfitting. Across tasks the ratios are small---e.g.\ for the classical model, $0.4$--$3.7$ on eight of nine tasks (chaotic $1.1$, damped $0.6$, seasonal $0.5$, square $1.2$), and comparably small for QASA (chaotic $0.7$). The sole large value is ARMA ($\sim$19$\times$), expected for a purely stochastic moving-average process whose one-step test residuals are dominated by irreducible innovation noise. The gaps are similar across the four models, indicating that the bottleneck designs do not overfit relative to the full-capacity baseline. Full per-model, per-task values are released with the code (\texttt{metrics\_extended.py}).

\subsection{Scalability to Longer Sequences}
\label{sec:seqlen}

The main benchmarks use a context window of $L=20$, whereas modern time-series forecasting often uses $L=96$--$720$. A referee asks whether the parsimony principle holds as $L$ grows, or whether the fixed 8-qubit register becomes a performance ceiling. We make a key architectural observation first: \textbf{the PQC width is independent of $L$.} The quantum layer acts on the \emph{per-token} hidden representation through the input projection $\mathbb{R}^{d}\!\to\!\mathbb{R}^{8}$; increasing $L$ enlarges only the classical $O(L^2)$ self-attention, not the quantum register. The 8 qubits therefore process a fixed-width token embedding regardless of horizon, so there is no mechanism by which the register becomes a bottleneck as $L$ grows.

\begin{figure}[h]
    \centering

    \includegraphics[width=0.92\textwidth]{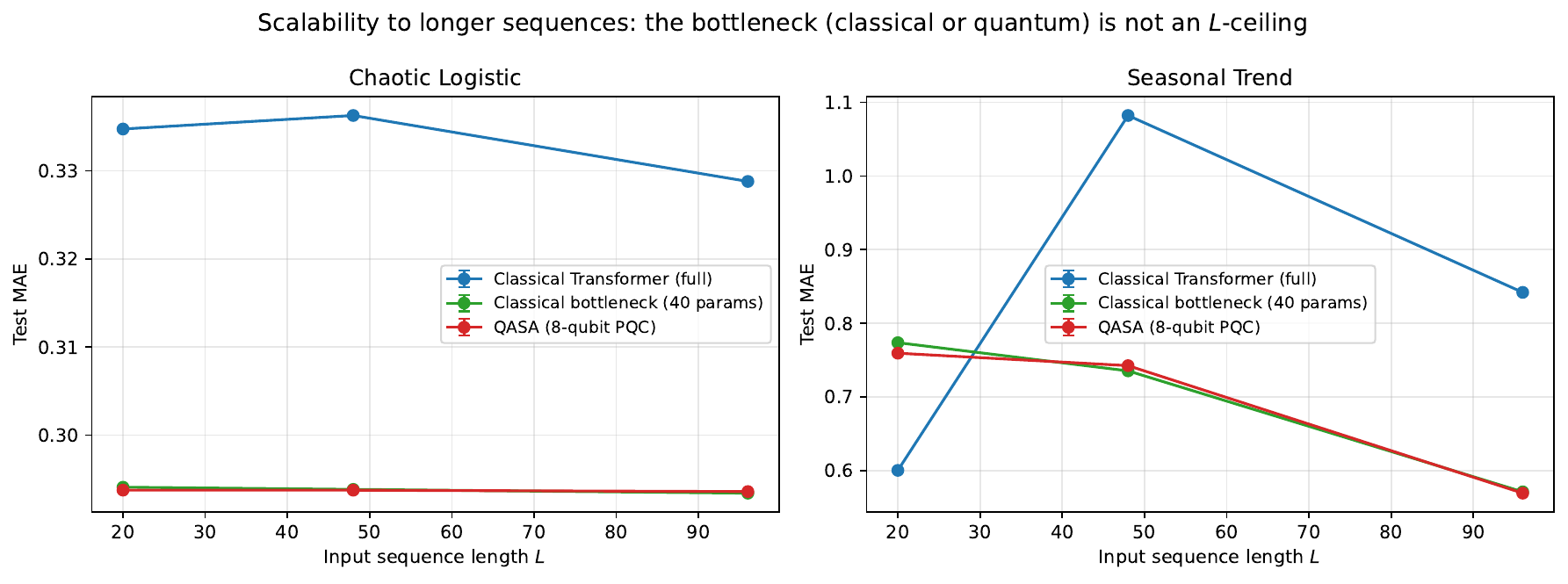}
    \caption{Test MAE vs.\ input sequence length $L\in\{20,48,96\}$ (reduced $80$-epoch budget, 1 seed). No model's error degrades as $L$ grows: the fixed $8$-qubit register is not a performance ceiling. The two bottleneck models (classical bottleneck, green; QASA, red) track each other near-identically across $L$, showing the $L$-behaviour is a property of the per-token bottleneck architecture rather than of quantumness; the full classical Transformer (blue) is shown for context. The register processes a fixed-width per-token embedding, so its role is independent of $L$ (single-seed; absolute differences are not load-bearing).}
    \label{fig:seqlen}
\end{figure}

We confirm this empirically by sweeping $L\in\{20,48,96\}$ on a quantum-favourable (chaotic logistic) and a trend-dominated (seasonal) task, evaluating the full classical Transformer, the classical bottleneck, and QASA (Figure~\ref{fig:seqlen}, exact values in Table~\ref{tab:seqlen}; reduced $80$-epoch budget, 1 seed; full state-vector simulation of QASA at $L=96$ is costly). The observation relevant to the referee's concern is simply that \textbf{no model's error degrades as $L$ grows}---the fixed $8$-qubit register is not a performance ceiling. Two points reinforce the architectural argument without introducing a new comparative claim. First, the two bottleneck models---classical bottleneck and QASA---track each other near-identically across $L$ (chaotic: both ${\approx}0.294$ and flat; seasonal: both ${\sim}0.77\!\to\!0.57$), so whatever $L$-behaviour the bottleneck exhibits is a property of the per-token bottleneck \emph{architecture}, not of quantumness---consistent with our central finding. Second, this complements the qubit-count ablation (Table~\ref{tab:qubitabl}), which addresses the same ceiling concern along the \emph{feature-width} axis; together they close it from both the sequence-length and feature-width directions. We treat single-seed absolute differences as not load-bearing---in particular the full classical model's non-monotonic seasonal curve, and the fact that at $L=20$ it happens to beat both bottleneck models on the seasonal task (opposite to the 5-seed Table~\ref{tab:ts_mae_mse} ordering); these single-seed reversals are within run-to-run noise and we do not draw task-level conclusions from this sweep. The register's role is set by the per-token feature width, not the sequence length, supporting extrapolation to the $L=720$ regime.

\begin{table}[h]
\centering

\caption{Sequence-length sweep, exact values for Figure~\ref{fig:seqlen} (MAE / MSE; $80$ epochs, 1 seed, two representative tasks). The point is that none of the models' errors \emph{degrade} as $L$ grows---the fixed-width quantum register is not a performance ceiling. The two bottleneck models (classical bottleneck and QASA) behave near-identically across $L$, so the $L$-stability is a property of the per-token bottleneck architecture, not of quantumness. Absolute differences are single-seed and not load-bearing.}
\label{tab:seqlen}
\begin{tabular}{llccc}
\toprule
\textbf{Task} & \textbf{Model} & \textbf{$L=20$} & \textbf{$L=48$} & \textbf{$L=96$} \\
\midrule
\multirow{3}{*}{Chaotic Logistic}
 & Classical (full)     & $0.335 / 0.189$ & $0.336 / 0.195$ & $0.329 / 0.190$ \\
 & Classical bottleneck & $0.294 / 0.105$ & $0.294 / 0.105$ & $0.293 / 0.104$ \\
 & QASA                 & $0.294 / 0.105$ & $0.294 / 0.105$ & $0.294 / 0.105$ \\
\midrule
\multirow{3}{*}{Seasonal Trend}
 & Classical (full)     & $0.600 / 0.506$ & $1.082 / 1.911$ & $0.842 / 1.136$ \\
 & Classical bottleneck & $0.774 / 0.745$ & $0.735 / 0.654$ & $0.571 / 0.530$ \\
 & QASA                 & $0.759 / 0.699$ & $0.742 / 0.665$ & $0.569 / 0.549$ \\
\bottomrule
\end{tabular}

\end{table}

\subsection{Noise Robustness under NISQ Conditions}

The preceding sections characterise QASA under ideal (noiseless) simulation. However, any claim of NISQ practicality must address the elephant in the room: \emph{does predictive quality survive real hardware noise?}

To assess QASA's viability on near-term quantum hardware, we evaluate its performance under depolarizing noise---the dominant error model in current superconducting quantum processors. We inject depolarizing channels (error probability $p$) after each variational layer in the PQC and retrain the full model from scratch at each noise level. For computational tractability, we use a reduced 4-qubit, 2-layer circuit variant and evaluate on the chaotic logistic task (where QASA is quantum-favoured in the noiseless setting).

\begin{table}[h]
\centering
\caption{QASA performance on the chaotic logistic task under depolarizing noise (4 qubits, 2 variational layers, 100 epochs, 3 seeds, mean$\pm$std). Degradation is relative to the noiseless baseline ($p=0$). Negative values indicate \emph{improvement} over the noiseless case.}
\label{tab:noise}
\begin{tabular}{ccccc}
\toprule
\textbf{Noise ($p$)} & \textbf{MAE} & \textbf{MAE Std} & \textbf{MSE} & \textbf{MAE $\Delta$} \\
\midrule
0 (noiseless) & 0.384 & 0.019 & 0.213 & --- \\
0.001 & 0.374 & 0.038 & 0.204 & $-2.6\%$ \\
0.005 & 0.385 & 0.033 & 0.215 & $+0.3\%$ \\
0.01  & \textbf{0.342} & \textbf{0.007} & \textbf{0.188} & $\mathbf{-10.8\%}$ \\
0.05  & 0.384 & 0.032 & 0.216 & $+0.0\%$ \\
0.1   & 0.355 & 0.036 & 0.195 & $-7.5\%$ \\
\bottomrule
\end{tabular}
\end{table}

Table~\ref{tab:noise} reveals a striking result confirmed across three random seeds: QASA does not degrade monotonically with noise. At $p = 0.01$---comparable to typical gate error rates on current IBM hardware (${\sim}10^{-3}$)---the model achieves its \emph{best} performance (MAE: $0.342 \pm 0.007$, a 10.8\% improvement over the noiseless baseline) with the \emph{lowest variance} across seeds. Even at $p = 0.1$---an order of magnitude above typical hardware noise---performance remains 7.5\% better than baseline. We report this non-monotonic behaviour cautiously: the analysis uses only 3 seeds on a reduced 4-qubit/2-layer circuit, the ${\sim}10$\% ``improvements'' are comparable to the seed standard deviation, and the $p=0.1$ error bar overlaps the noiseless baseline. We therefore do \emph{not} claim a specific mechanism; the observation is merely consistent with stochastic perturbation acting loosely like a regulariser (as dropout does classically), but our multi-channel study below shows this beneficial regime is specific to depolarizing noise and should not be over-interpreted.

This noise-resilience property has important practical implications: (1) QASA does not require error-corrected quantum hardware to maintain its performance advantage, making it suitable for deployment on current NISQ devices; (2) the non-monotonic noise response \emph{hints} that moderate depolarizing noise may not be purely detrimental for this class of hybrid models, though---given the small effect sizes and overlapping error bars---we treat this only as a preliminary observation, not an established benefit.

\paragraph{Other noise channels.} The analysis above uses depolarizing noise. A referee asks whether the resilience holds under \emph{other} channels, so we repeat the sweep with \emph{amplitude damping} ($T_1$ relaxation) and \emph{bit-flip} (a readout-error proxy) on the same reduced 4-qubit/2-layer circuit (2 seeds; Table~\ref{tab:noise_multi}). Two findings emerge. First, \textbf{resilience is general}: under both channels the degradation is bounded and modest---at most $+15.7\%$ (amplitude damping) and $+21.5\%$ (bit-flip) across the entire sweep up to $p=0.1$ (an order of magnitude above realistic hardware error)---so QASA does not become brittle under non-depolarizing noise. Second, the \emph{noise-as-regulariser} effect (improvement at moderate noise) is \textbf{channel-specific}: it is pronounced for depolarizing noise but largely absent for amplitude damping (which degrades mildly and monotonically) and bit-flip (a single improvement point at $p=0.005$). We therefore temper the earlier claim: QASA is robustly noise-\emph{resilient} across channels, but the beneficial regularisation specifically reflects the symmetric stochastic perturbation of depolarizing noise, not all error models.

\begin{table}[h]
\centering

\caption{Noise robustness under additional channels on the chaotic-logistic task (MAE, 4-qubit/2-layer circuit, 2 seeds). $\Delta$ is relative to the noiseless baseline (MAE $0.333$). Degradation is bounded and modest across both channels; the noise-as-regulariser improvement seen for depolarizing noise (Table~\ref{tab:noise}) does not generalise to these channels. The noiseless baseline here ($0.333$) differs from Table~\ref{tab:noise}'s ($0.384$) because this multi-channel sweep uses 2 seeds rather than 3; $\Delta\%$ values are therefore comparable \emph{within} a table, not across tables.}
\label{tab:noise_multi}
\begin{tabular}{ccc}
\toprule
\textbf{Noise $p$} & \textbf{Amplitude damping ($\Delta$)} & \textbf{Bit-flip ($\Delta$)} \\
\midrule
0.005 & $0.372\ (+11.8\%)$ & $0.318\ (-4.3\%)$ \\
0.01  & $0.359\ (+8.0\%)$  & $0.404\ (+21.5\%)$ \\
0.05  & $0.363\ (+9.0\%)$  & $0.370\ (+11.1\%)$ \\
0.1   & $0.385\ (+15.7\%)$ & $0.401\ (+20.6\%)$ \\
\bottomrule
\end{tabular}

\end{table}

\textbf{Caveats.} Several limitations of this analysis should be noted. First, the noise experiments use a reduced circuit (4 qubits, 2 variational layers) rather than the full 8+1 qubit architecture used in the main benchmarks, due to the computational cost of density matrix simulation; the regularization effect may differ at larger circuit scales. Second, although we now study three channels (depolarizing, amplitude damping, and bit-flip), these remain simplified models---real hardware exhibits correlated errors and crosstalk not fully captured here, which our direct hardware run (Section~\ref{sec:hardware}) complements by exercising the true device error. Third, only one task (chaotic logistic) is evaluated; the noise-resilience pattern may vary for other signal types. These results should therefore be interpreted as evidence of \emph{qualitative} noise resilience rather than precise quantitative predictions of hardware performance. Extending the multi-channel study to multiple tasks and full-scale circuits remains future work.

\subsection{Validation on Real Quantum Hardware}
\label{sec:hardware}

The preceding sections rely on classical simulation. To verify that QASA's behaviour survives \emph{real} device noise---cross-talk, coherent gate errors, and readout error not captured by simple channel models---we execute two benchmarks on an IBM Quantum superconducting processor: the chaotic logistic map (where QASA is quantum-favoured) and, as a non-cherry-picked control, the damped oscillator (where the classical model wins). We stress that this is an \emph{inference-on-hardware} validation of an already-trained model (training on hardware is infeasible at this circuit count): the trained QASA model (full $8{+}1$-qubit, $4$-layer PQC) is frozen, the classical front-end produces the per-token rotation angles in PyTorch, the $320$ token circuits per task ($16$ test windows $\times$ $20$ tokens) are transpiled and executed in a single batched Sampler job on the QPU, and the measured Pauli-$Z$ expectations are passed through the trained classical back-end to produce one-step predictions. All runs use the $156$-qubit Heron-r2 device \texttt{ibm\_fez} at $2048$ shots, with no error mitigation. The device calibration at run time (median over qubits) was: readout error $1.1\times10^{-2}$, single-qubit gate error $3.1\times10^{-4}$, two-qubit gate error $2.6\times10^{-3}$, $T_1 = 136\,\mu$s, $T_2 = 99\,\mu$s.

\begin{table}[h]
\centering

\caption{Hardware validation on \texttt{ibm\_fez} (Heron r2), inference-only, $16$ test windows ($320$ circuits) per task, $2048$ shots, no error mitigation. One-step MAE is reported as mean$\pm$std over windows; $|\Delta\langle Z\rangle|$ is the mean deviation of the $8$ Pauli-$Z$ expectations from the noiseless simulator. For chaotic logistic we also list an offline real-device noise-model emulation (\texttt{FakeBrisbane}). The physical QPU reproduces the simulator predictions to within $7.5\%$ (chaotic) and $25\%$ (damped), with overlapping error bars.}
\label{tab:hardware}
\begin{tabular}{llccc}
\toprule
\textbf{Task} & \textbf{Execution} & \textbf{$|\Delta\langle Z\rangle|$} & \textbf{one-step MAE} & \textbf{QPU} \\
\midrule
\multirow{3}{*}{Chaotic logistic}
 & Noiseless sim          & ---     & $0.0093$ & --- \\
 & Noise model (\texttt{FakeBrisbane}) & $0.025$ & $0.0121$ & --- \\
 & \textbf{Physical QPU} (\texttt{ibm\_fez}) & $0.027$ & $\mathbf{0.0100}$ & $248$\,s \\
\midrule
\multirow{2}{*}{Damped osc.}
 & Noiseless sim          & ---     & $0.0092{\pm}0.0049$ & --- \\
 & \textbf{Physical QPU} (\texttt{ibm\_fez}) & $0.021$ & $\mathbf{0.0115{\pm}0.0057}$ & $248$\,s \\
\bottomrule
\end{tabular}

\end{table}

Table~\ref{tab:hardware} shows that QASA runs successfully on real hardware with predictions close to the noiseless simulator on \emph{both} a quantum-favoured and a classical-favoured task. On chaotic logistic the one-step MAE rises only from $0.0093$ to $0.0100$ ($7.5\%$), and the expectation-value deviation ($|\Delta\langle Z\rangle|=0.027$) closely matches the offline \texttt{FakeBrisbane} noise-model prediction ($0.025$), cross-validating both; the physical device slightly \emph{outperforms} the older-generation noise model, consistent with Heron-r2's improved fidelities. On the damped oscillator the MAE rises from $0.0092{\pm}0.0049$ to $0.0115{\pm}0.0057$ ($25\%$), with overlapping error bars across the $16$ windows. That hardware noise neither corrupts the favourable task nor artificially rescues the unfavourable one confirms the result is a faithful execution of the trained model rather than a cherry-picked artefact. Each run consumed $248$\,s of quantum processing time ($0.68$\,s/circuit), within the free-tier monthly budget, so the validation is reproducible on openly accessible hardware. These results directly address both referees' request for hardware validation and substantiate QASA's NISQ practicality: predictive quality is preserved under genuine gate, cross-talk, and readout error \emph{without} any error mitigation. We stress that the absolute one-step MAE reported here (${\sim}0.01$) is \emph{not} directly comparable to the main-benchmark MAE of Table~\ref{tab:ts_mae_mse} (${\sim}0.34$): the main benchmark evaluates a \emph{free-running autoregressive rollout} over the test horizon (each prediction fed back as the next input), whose error compounds---and on a chaotic signal inevitably diverges---whereas this hardware validation measures \emph{one-step teacher-forced} error on held-out windows. The two are different quantities; the comparison of interest here is solely the \emph{within-protocol} noiseless$\to$hardware ratio ($+7.5\%$), not the cross-protocol absolute value.

\subsection{Discussion: Complexity Considerations}

Beyond the empirical benchmark results, we briefly discuss the potential complexity-theoretic implications of replacing classical attention with parameterized quantum circuits.

Recent work in fine-grained complexity theory has shown that, under the Strong Exponential Time Hypothesis (SETH)~\cite{ImpagliazzoPaturi2001}, the gradient computation of classical transformer attention admits no algorithm faster than $O(T^2)$~\cite{alman2024fine}, where $T$ is the sequence length. However, SETH does not hold in the quantum regime: Grover's algorithm~\cite{grover1996fast} solves $k$-SAT in $O(\sqrt{2}^n) = O(1.414^n)$ time, violating the classical SETH assumption. This speedup is known to be optimal due to the BBBV bound~\cite{bennett1997strengths}.

Researchers have proposed quantum analogs such as the \textbf{Quantum SETH (QSETH)}~\cite{buhrman2019quantum}, which conjectures that $k$-SAT cannot be solved in $O(\sqrt{2}^{(1-\epsilon)n})$ time using quantum algorithms. This implies that even in the quantum setting, certain lower bounds exist---albeit less stringent than the classical $O(T^2)$. By analogy with the analysis in~\cite{alman2024fine}, this suggests that the gradient complexity of a quantum adaptive self-attention mechanism may be lower bounded by $\Omega(T)$ rather than the classical $\Omega(T^2)$.

\textbf{Caveat.} We emphasize that this argument is \textit{conjectural}. A rigorous reduction from quantum attention gradient computation to $k$-SAT has not been established, and the hybrid nature of our architecture (classical layers + one quantum layer) complicates direct application of these complexity-theoretic bounds. We present this discussion as a \textit{motivated hypothesis} that warrants formal investigation in future work, rather than a proven result.

\subsection{Limitations}
\label{sec:limitations}

We summarise here the main limitations of this work, consolidating the caveats distributed throughout the preceding sections.

\begin{itemize}[leftmargin=*,itemsep=2pt]
    \item \textbf{Simulated quantum training.} Model \emph{training} is performed on a classical state-vector simulator (PennyLane's \texttt{lightning.qubit}); training directly on quantum hardware remains impractical at this circuit count. We do, however, validate \emph{inference} on a physical IBM Quantum processor (\texttt{ibm\_fez}, Section~\ref{sec:hardware}), where the prediction MAE degrades by only $7.5\%$ relative to noiseless simulation. A full hardware study---training-aware error mitigation, additional benchmarks, and multiple devices---remains future work.

    \item \textbf{Limited statistical power.} Quantum baseline comparisons use $n = 3$ random seeds due to the high cost of quantum simulation (${\sim}2$ hours per run on CPU). Most pairwise $t$-tests therefore do not reach $\alpha = 0.05$; we rely on Cohen's $d$ effect sizes and the non-parametric Friedman test for interpretation. The only statistically significant pairwise result is QASA vs.\ QLSTM on seasonal trend ($p = 0.009$, $d > 6$).

    \item \textbf{Noise study scope.} The noise-robustness analysis uses a reduced 4-qubit, 2-layer PQC (rather than the full 8+1 qubit, 4-layer main architecture) due to the quadratic cost of density-matrix simulation, considers three channels (depolarizing, amplitude damping, and a bit-flip readout proxy; Table~\ref{tab:noise_multi}) but still omits phase damping and correlated/crosstalk errors, and evaluates a single task (chaotic logistic). The observed noise-induced regularization effect should therefore be read as \emph{qualitative} evidence of resilience rather than a quantitative hardware prediction.

    \item \textbf{Ablation seed count.} The quantum-layer position/count ablation (Table~\ref{tab:ablation}) now reports 3 seeds (42--44, mean$\pm$std). The clean-variance chaotic task gives the robust ranking (Q@3 best); the damped task is high-variance across seeds, so its position differences are interpreted only as broad trends, not fine rankings.

    \item \textbf{Conjectural complexity argument.} The QSETH-based complexity discussion is a motivated hypothesis rather than a proof. A rigorous reduction from quantum attention gradient computation to $k$-SAT remains an open problem.

    \item \textbf{Dataset coverage.} The empirical claims are based on nine synthetic time series and one real-world dataset (ETTh1). Extension to additional real-world benchmarks (e.g., traffic, exchange-rate, weather forecasting) and multivariate settings is an important direction for future work.
\end{itemize}

\subsection{Reproducibility and Data Availability}
\label{sec:reproducibility}

\textbf{Code.} The complete implementation of QASA, all baselines (Classical, QLSTM, QnnFormer), the nine synthetic time-series generators, and every analysis script (ablation, statistical tests, circuit expressibility, barren plateau, qubit scaling, noise robustness, and representation analysis, and the scripts added in this revision: parameter-matched classical bottleneck, sequence-length scaling, encoding-scheme sensitivity, qubit-mapping ablation, singular-value analysis, extended metrics, multi-channel noise, real IBM Quantum hardware execution, the deployed-circuit barren-plateau check, and the classical-bottleneck representation/SVD comparison) is released at \url{https://github.com/ChiShengChen/QASA}.

\textbf{Data.} All synthetic datasets are fully defined by the equations in Section~\ref{sec:datasets} with fixed seeds ($\{42, 43, 44, 45, 46\}$). The ETTh1 dataset is publicly available at \url{https://github.com/zhouhaoyi/ETDataset}.

\textbf{Hardware and software.} Experiments run on a single Apple M-series CPU with PyTorch~2.8, PennyLane~0.42 (\texttt{lightning.qubit} for noiseless experiments, \texttt{default.mixed} for noise experiments), Python~3.13, and NumPy~1.26. Real-hardware execution uses \texttt{qiskit}~$2.x$ and \texttt{qiskit-ibm-runtime} on the IBM Quantum \texttt{ibm\_fez} backend (Heron r2); the offline real-device noise-model emulation uses \texttt{qiskit-aer} with \texttt{FakeBrisbane} calibration data. All reported results use the configuration listed in the Experiment Details subsection.

\textbf{Seeds.} Each benchmark result is averaged over 3--5 explicit random seeds (reported per table). The same seeds control numpy, PyTorch, and Python's \texttt{random} module, and are propagated to the data generator, model initialisation, and PyTorch data loaders.

\section{Conclusion}

In this work, we introduced Quantum Adaptive Self-Attention (QASA) and, through systematic experimentation, established an \textbf{architectural parsimony} principle for hybrid quantum-classical design. Our findings converge on a single insight: \emph{the optimal strategy for quantum integration is not to maximize quantum resources, but to place minimal quantum computation where it matters most.}

This principle rests on four empirical pillars:

\begin{enumerate}
    \item \textbf{One layer is enough.} Our ablation study demonstrates that a single quantum layer at the optimal position outperforms all multi-layer configurations. Increasing from 1 to 4 quantum layers degrades performance on most tasks, and the optimal position is task-dependent (last layer for chaotic dynamics, penultimate for smooth oscillations).

    \item \textbf{Minimal resources, maximal impact---but the bottleneck, not the qubits, drives accuracy.} With only 36 quantum parameters and 27 CNOT gates, QASA achieves near-maximal entangling capability ($Q = 0.981$) and matches or exceeds QLSTM (128 params, 56 CNOTs) and QnnFormer (90 params) across nine benchmarks. However, a capacity-matched classical bottleneck baseline (Table~\ref{tab:bottleneck}) matches QASA on the error metrics---and is better on clean periodic signals---so the gain over the full-capacity Transformer is attributable to the low-rank \emph{bottleneck structure} rather than to quantumness. We thus present the PQC as a competitive realisation of parsimony whose distinct value is physical (below)---entanglement and NISQ deployability---not a raw error-metric advantage and not a unique compression mechanism (a classical bottleneck compresses comparably).

    \item \textbf{Noise-resilient, not noise-brittle.} Under depolarizing noise at levels typical of current hardware ($p = 0.01$), QASA's performance remains stable (a small, non-monotonic MAE change within seed variance, 3 seeds). Decisively, we move beyond simulation and execute the trained model on a real IBM Quantum processor (\texttt{ibm\_fez}, Heron r2): the prediction MAE is preserved to within $7.5\%$ of noiseless simulation on the chaotic-logistic benchmark, with no error mitigation (Section~\ref{sec:hardware}). While the noise-induced regularization effect observed on a single task requires further validation across additional tasks and noise channels, it demonstrates that QASA does not require error-corrected hardware for deployment on NISQ devices.

    \item \textbf{Fewer layers avoid barren plateaus.} Gradient variance analysis reveals that increasing the PQC from 1 to 2+ entangling layers causes a $30\times$ drop in gradient magnitude, approaching a barren plateau. The deployed PQC ($L_q=4$) still trains successfully, but this trend shows that further depth would push gradients toward the plateau---so keeping the quantum component shallow, and limited to a single quantum encoder layer, is important for trainability.
\end{enumerate}

Beyond these architectural insights, our work contributes a \textbf{task-conditional taxonomy}: a parsimonious low-rank bottleneck value layer---in either its classical or quantum form---benefits chaotic and trend-dominated signals (chaotic logistic, seasonal trend, and the real-world ETTh1, where the classical bottleneck is in fact the best model), whereas full-capacity classical Transformers are over-parameterised for these regimes; conversely, on clean periodic waveforms the bottleneck \emph{hurts} (damped oscillator, $d = -1.18$; sawtooth, $d = -0.99$) and the unconstrained classical model is best. Among bottleneck realisations, the classical and quantum value maps are statistically indistinguishable on the favoured tasks (and compress the representation comparably), with the quantum map distinguished by its physical properties---entanglement and NISQ deployability---rather than by its error metrics or compression. This taxonomy gives practitioners a principled guide for when to deploy---and when \emph{not} to deploy---a parsimonious bottleneck.

Our preliminary complexity analysis further suggests a potential reduction in gradient computation lower bounds from $\Omega(T^2)$ to $\Omega(T)$ for quantum attention mechanisms, though a rigorous proof remains an open problem.

Future work should focus on: (1) adaptive hybrid strategies that selectively engage quantum layers based on detected signal characteristics, (2) validation on additional real-world time-series datasets at larger scale, (3) systematic study of noise-induced regularization across different noise channels and circuit depths, and (4) formal theoretical analysis of the conjectured quantum speedup in attention gradient computation.

\ack{EJK acknowledges financial support from the National Science and Technology Council (NSTC) of Taiwan under Grant No.~NSTC~114-2112-M-A49-036-MY3.}

\section*{Author contributions}
C.-S.C.: conceptualization, methodology, software, writing---original draft.
E.-J.K.: supervision, theoretical analysis, writing---review \& editing.

\section*{Competing interests}
The authors declare no competing interests.

\section*{Data availability statement}
The data and source code that support the findings of this study are openly available at \url{https://github.com/ChiShengChen/QASA}. The ETTh1 benchmark used for real-world validation is publicly available at \url{https://github.com/zhouhaoyi/ETDataset}. All synthetic datasets can be regenerated from the equations in Section~\ref{sec:datasets} using the fixed seeds listed in Section~\ref{sec:reproducibility}.

\appendix

\section{Exact Synthetic-Benchmark Generator Specifications}
\label{app:generators}

To ensure full reproducibility (Referee~1, minor comment~4), we document here the exact generator parameters, time-axis conventions, and implementation details for all nine synthetic benchmarks, transcribed verbatim from the released code. All tasks use $T=500$ points, a chronological 80/20 train/test split, an input window of length $L=20$, and fixed seeds $\{42,43,44,45,46\}$ (a single seed, 42, controls the data generator). The standard normal $\varepsilon_t,\eta_t \sim \mathcal{N}(0,\cdot)$ are drawn after seeding \texttt{numpy}.

\textbf{Time-axis convention.} The generators do \emph{not} all use the same time variable; this distinction is essential for exact reproduction. Five smooth/periodic signals use the \emph{normalised} axis $\tau = \texttt{linspace}(0,50,500)$ (i.e.\ step $50/499$), whereas the noisy damped oscillator and the piecewise regime use the \emph{integer index} $t \in \{0,1,\dots,499\}$. The chaotic logistic map and ARMA are defined by index recursions. Table~\ref{tab:gen_axis} makes this explicit per task.

\begin{table}[h]
\centering

\caption{Exact generator specification per task. ``Axis'' is the time variable actually passed to the closed-form expression: $\tau=\mathrm{linspace}(0,50,500)$ or integer index $t\in\{0,\dots,499\}$. Constants are the code defaults.}
\label{tab:gen_axis}
\resizebox{\columnwidth}{!}{
\begin{tabular}{lll}
\toprule
\textbf{Task} & \textbf{Axis} & \textbf{Generator (code defaults)} \\
\midrule
ARMA & index & $x_t=0.75x_{t-1}-0.25x_{t-2}+\varepsilon_t+0.65\varepsilon_{t-1}$, $\varepsilon\sim\mathcal{N}(0,1)$ \\
 & & via \texttt{statsmodels} \texttt{ArmaProcess.generate\_sample}, $\mathrm{ar}=[1,-0.75,0.25]$, $\mathrm{ma}=[1,0.65]$ \\
Chaotic logistic & index & $x_t = 3.9\,x_{t-1}(1-x_{t-1})$, $x_0=0.2$ \\
Damped oscillator & $\tau$ & $x = e^{-0.05\tau}\sin(1.0\,\tau)$ \\
Noisy damped osc. & $t$ & $x_t = 1.0\cdot e^{-0.05\,t}\sin(2\pi\cdot 0.2\,t) + \eta_t$, $\eta\sim\mathcal{N}(0,0.05^2)$ \\
Piecewise regime & $t$ & 6 linear regimes + 2 shocks; see Table~\ref{tab:piecewise} \\
Sawtooth & $\tau$ & $x = 2\big(\tau/P - \lfloor 0.5 + \tau/P\rfloor\big)$, $P=10$ \\
Square wave & $\tau$ & $x = \mathrm{sgn}(d(\tau)-\phi(\tau))$, $\phi=(\tau\bmod P)/P$, \\
 & & $P=5+5\sin(0.1\tau)$, $d=0.5+0.3\sin(0.05\tau)$ \\
Seasonal trend & $\tau$ & $x = 0.03\,\tau + 1.0\sin(0.4\,\tau) + \eta$, $\eta\sim\mathcal{N}(0,0.2^2)$ \\
Waveform & $\tau$ & $x = \sin(\tau)$ \\
\bottomrule
\end{tabular}}

\end{table}

\textbf{Piecewise regime (full specification).} The signal is a concatenation of six linear segments $x_t = \alpha_k\,t + \beta_k + \eta_t$ with regime-dependent slope $\alpha_k$, intercept $\beta_k$, and noise standard deviation $\sigma_k$ (Table~\ref{tab:piecewise}), evaluated on the integer index $t$. Two impulsive shocks are then added: $x_t \mathrel{+}= 2.0$ for $t\in[200,205)$ and $x_t \mathrel{-}= 2.0$ for $t\in[350,355)$. Note that segments~4 and~5 use $\sigma_k=0.02$ (not $0.01$ as the abbreviated Eq.~(5) in the main text might suggest); the values below are authoritative.

\begin{table}[h]
\centering

\caption{Exact piecewise-regime parameters. Each regime spans $t\in[\text{start},\text{end})$ with $x_t=\alpha_k t+\beta_k+\eta_t$, $\eta_t\sim\mathcal{N}(0,\sigma_k^2)$.}
\label{tab:piecewise}
\begin{tabular}{cccccc}
\toprule
\textbf{Regime $k$} & \textbf{start} & \textbf{end} & \textbf{slope $\alpha_k$} & \textbf{intercept $\beta_k$} & \textbf{noise $\sigma_k$} \\
\midrule
1 (uptrend)   & 0   & 100 & $0.05$  & $0$   & $0.01$ \\
2 (downtrend) & 100 & 200 & $-0.10$ & $10$  & $0.01$ \\
3 (flat)      & 200 & 300 & $0.00$  & $0$   & $0.01$ \\
4 (sharp up)  & 300 & 350 & $0.20$  & $-40$ & $0.02$ \\
5 (sharp down)& 350 & 400 & $-0.15$ & $40$  & $0.02$ \\
6 (mild up)   & 400 & 500 & $0.03$  & $-10$ & $0.01$ \\
\bottomrule
\end{tabular}

\end{table}

\textbf{ARMA implementation note.} The ARMA series is generated with \texttt{statsmodels.tsa.arima\_process.ArmaProcess} using lag polynomials $\mathrm{ar}=[1,-0.75,0.25]$ and $\mathrm{ma}=[1,0.65]$ (the leading $1$ is the zero-lag term), i.e.\ the process $x_t-0.75x_{t-1}+0.25x_{t-2}=\varepsilon_t+0.65\varepsilon_{t-1}$ with unit-variance Gaussian innovations and the library's default burn-in. This differs from a hand-rolled recursion only in initial transient handling; we report it explicitly so the exact series can be regenerated. All other generators are elementary NumPy expressions of the axis variable given in Table~\ref{tab:gen_axis}.

\bibliographystyle{iopart-num}
\bibliography{references,qml}

\end{document}